\documentclass[prd, twocolumn, nofootinbib, superscriptaddress,physrev,aps]{revtex4-2}

\usepackage{amsmath,amssymb,physics}
\usepackage[amsmath,thmmarks]{ntheorem}

\usepackage{newtxtext}
\usepackage[scaled]{helvet}
\usepackage{bm} 
\usepackage{graphicx}
\usepackage{dcolumn} 
\usepackage[dvipsnames]{xcolor}
\usepackage{colortbl} 
\usepackage{rotating} 
\usepackage[setpagesize=false, bookmarks=true, bookmarksdepth=tocdepth, bookmarksnumbered=true, colorlinks=true, citecolor=MidnightBlue, linkcolor=MidnightBlue, pdftitle={}, pdfsubject={}, pdfauthor={}, pdfkeywords={}]{hyperref}

\begin{document}

\title{Constraining the lensing dispersion from the angular clustering of binary black hole mergers}

\author{Fumihiro Chuman}
\affiliation{Department of Physics, Graduate School of Science, Chiba University, 1-33 Yayoicho, Inage, Chiba 263-8522, Japan}

\author{Masamune Oguri}
\affiliation{Center for Frontier Science, Chiba University, 1-33 Yayoicho, Inage, Chiba 263-8522, Japan}
\affiliation{Department of Physics, Graduate School of Science, Chiba University, 1-33 Yayoicho, Inage, Chiba 263-8522, Japan}

\date{\today}

\begin{abstract}
Gravitational waves from inspiraling compact binaries provide direct measurements of luminosity distances and serve as a powerful probe of the high-redshift Universe. In addition to their role as standard sirens, they offer an opportunity to constrain small-scale density fluctuations through the dispersion in the distance-redshift relation induced by gravitational lensing.
We propose a method to constrain this lensing dispersion without requiring the redshift information by analyzing the angular clustering of gravitational wave sources. Our formalism incorporating second-order lensing effects in the luminosity distance shows that the amplitude of the auto-correlation angular clustering decreases with increasing lensing dispersion. While we show that the auto-correlation signal is detected with sufficient signal-to-noise ratios in future gravitational wave experiments, there exists a strong degeneracy between the lensing dispersion and the linear bias of gravitational wave sources. We demonstrate that this degeneracy is partially broken by a joint analysis of the auto-correlation of gravitational wave sources and the cross-correlation with galaxies whose redshifts are known. This approach enhances the use of gravitational waves as a cosmological probe at high redshifts.

\end{abstract}

\maketitle

\section{Introduction}
Gravitational lensing due to cosmic inhomogeneities induces the dispersion in the distance-redshift relation. This dispersion can constrain small-scale density fluctuations and has been proposed as a useful cosmological probe in several ways. For instance, it can be used to constrain the abundance of primordial black holes (e.g., \citep{zumalacarregui2018,shah2025a}), place an upper limit on neutrino masses (e.g., \citep{hada2016, agrawal2019c}), probe the small-scale cosmological density power spectrum (e.g., \citep{ben-dayan_constraints_2016}), and constrain the primordial power spectrum (e.g., \citep{ben-dayan2014}). 

In these studies, Type Ia supernovae are considered as a probe of the dispersion in the distance-redshift relation. However, constructing a large sample of Type Ia supernovae at high redshifts is challenging as they require wide-field cadenced surveys in near-infrared (e.g., \citep{galbany2023}).

Gravitational waves from inspiraling  compact binaries offer an alternative means of measuring the distance-redshift relation because luminosity distances to the binaries can be directly measured from gravitational waveforms \citep{schutz1986}. These waveforms can be accurately predicted within the framework of general relativity. They can be observed with high precision through the matched filtering analysis, allowing accurate and precise measurements of luminosity distances. For this reason, inspiraling  compact binaries are often referred to as standard sirens, analogous to standard candles for Type Ia supernovae. For instance, \citet{holz2005} argue that luminosity distances derived from gravitational waveforms of binary black holes can be measured with an accuracy of less than $10\%$ even at high redshifts (e.g., $z\sim3$), independently of electromagnetic observations. Thus, standard sirens, particularly binary black hole mergers, have the potential to enable precise measurements of the lensing dispersion of the distance-redshift relation at high redshifts, which is challenging with Type Ia supernova observations.

However, a challenge lies in the fact that binary black holes tend to lack redshift information because binary black hole mergers are not expected to produce electromagnetic emissions. Electromagnetic counterparts may occur only in unique environmental conditions \citep{loeb2016}. Furthermore, the poor localization accuracy of gravitational wave sources \citep{abbott2020} prevents us from identifying their host galaxies. As such, binary black holes are sometimes called dark sirens due to their lack of redshift information.
In contrast, bright sirens, such as binary neutron stars and binary black holes accompanied by transient phenomena, provide redshift information via observations of electromagnetic counterparts and hence allow us to directly constrain the luminosity-distance relation and its dispersion caused by gravitational lensing (e.g., \citep{cutler2009a, gupta2019}). Given that most of the gravitational waves observed so far are dark sirens \citep{collaboration2023}, it is crucial to develop methods of utilizing dark sirens for obtaining cosmological information. 

Indeed, various techniques have been proposed to extract cosmological information, particularly the Hubble constant $H_0$, from dark sirens without electromagnetic counterparts. For example, by combining the luminosity distance and the cosmological phase shift, the expansion history of the Universe can be measured without any reference to the electromagnetic counterpart or the host-galaxy identification (e.g., \citep{nishizawa2012}). Another approach to constrain $H_0$ using gravitational-wave observations alone exploits  the narrow mass distribution of neutron stars to break the redshift–chirp mass degeneracy (e.g., \citep{taylor2012, taylor2012a}). Tidal effects in neutron star mergers can also break the degeneracy between the redshift and the chirp mass, enabling redshift estimation from gravitational-wave signals alone (e.g., \citep{messenger2012}).
Utilizing the anisotropies of the number density and luminosity distances of compact binaries originating from the large-scale structure, tight constraints on primordial non-Gaussianity can also be obtained without redshift information (e.g., \citep{namikawa2016}). Another approach uses the shape of the black hole mass function to constrain the source mass and redshift (e.g., \citep{mastrogiovanni2021,ezquiaga2022}). One of the most widely adopted approaches to dark sirens to date is the statistical assignment of host galaxies to gravitational wave sources using the galaxy catalogs for constraining $H_0$ (e.g., \citep{schutz1986, macleod2008, pozzo2012, kyutoku2017, thedescollaboration2019, gray2020, theligoscientificcollaboration2021}). If a host galaxy can be assigned, the redshift of the gravitational wave source can be inferred from the redshift of the host galaxy. The constraining power of this method can be improved as the number of gravitational wave sources increases. However, the accuracy of this method depends critically on the quality of galaxy catalogs, which is often difficult to obtain (e.g., \citep{borhanian2020,muttoni2023,dalang2024}). 

A promising and robust alternative approach that is less affected by uncertainties of the galaxy catalog is to perform a statistical analysis based on the spatial cross-correlation between the distribution of binary black holes and that of galaxies (e.g., \citep{oguri2016,mukherjee2021, mukherjee2024}). Recently, this method has been applied to realistic mock datasets of gravitational wave sources to demonstrate its robustness and feasibility \citep{ferri2025, pedrotti2025}. This approach is insensitive to the modeling of the binary black hole population, the merger rate, and the linear bias, which is an advantage of this method.

In this paper, we propose to utilize the angular clustering of gravitational wave sources to constrain the dispersion of the distance-redshift relation, which contains information on the gravitational lensing convergence. Previous studies using the spatial cross-correlation between binary black holes and galaxies have incorporated lensing-induced biases in the luminosity distance (e.g., \citep{oguri2016}), and in some cases have incorporated higher-order lensing effects into the observational uncertainties. Moreover, some previous studies have attempted to probe the gravitational lensing bias in this context (e.g., \citep{scelfo2018,scelfo2020,begnoni2024a}). 
However, to date, the impact of the lensing \textit{dispersion} arising from the inhomogeneous matter distribution along the line of sight on the angular clustering has not been systematically investigated. We show how clustering observables such as the auto-correlation angular power spectrum of binary black holes on the celestial sphere and the cross-correlation angular power spectrum between binary black holes and galaxies with known redshifts depend on the lensing dispersion and, hence, the lensing convergence. We compute the cumulative signal-to-noise ratio and conduct the Fisher matrix analysis to discuss the detectability. We pay particular attention to the degeneracy between the lensing dispersion and the linear bias of gravitational wave sources. 

This paper is organized as follows. In Sec.~\ref{sec:method}, we formulate angular clustering signals and the analysis method. We present our results, including the cumulative signal-to-noise ratio and the Fisher matrix analysis in Sec.~\ref{sec:result}. We also discuss the impact of the binary black hole merger rate, the choice of luminosity distance and redshift bin widths, and different treatments of the lensing-induced dispersion in Sec.~\ref{sec:discussion}. Finally, we present our conclusion in Sec.~\ref{sec:conclusion}.
Throughout the paper, we assume a flat Friedmann-Lemaître-Robertson-Walker (FLRW) universe as our cosmological model, with cosmological parameter values based on the latest Planck observations \cite{collaborationPlanck2018Results2020}.
The fiducial parameter values are as follows: the matter density parameter $\Omega_{\mathrm{m}}=0.3111$, the dark energy density parameter $\Omega_{\mathrm{de}}=0.6888$, the baryon density parameter $\Omega_{\mathrm{b}}=0.0490$, the dimensionless Hubble constant $h=0.6766$, the spectral index $n_s=0.9665$, the amplitude of density fluctuations smoothed at the $8\;\mathrm{Mpc}/h$ scale $\sigma_8=0.8102$, and the dark energy equation of state $w_{\mathrm{de}}=-1$. We used Colossus~\citep{diemer_colossus_2018}, a Python package for cosmological and large-scale structure calculations, to calculate cosmological distances and the linear matter power spectrum.

\section{Method}\label{sec:method}
We begin by deriving an expression for the luminosity distance that accounts for both gravitational lensing and measurement errors. This expression forms the basis for constructing a selection function, which we use to define a luminosity distance bin for the source distribution.
Using the selection function, we project the three-dimensional number density field of gravitational wave sources onto the two-dimensional celestial sphere. From the projected number density, we compute the two-dimensional number density fluctuations by subtracting the mean. The same procedure is applied to spectroscopic galaxies. Using these fluctuations, we calculate the auto-correlation angular power spectrum of gravitational wave sources and the cross-correlation angular power spectrum between gravitational wave sources and spectroscopic galaxies. These angular power spectra are evaluated as a function of the dispersion of the lensing convergence, which is treated as a free parameter. We present expressions for the signal-to-noise ratios and the Fisher matrix, which are used to assess the feasibility.

\subsection{Luminosity Distance to Gravitational Wave Sources}\label{subsec:luminosity distance}
While we cannot directly measure redshifts of gravitational wave sources from observed gravitational wave signals under usual circumstances, the waveform analysis allows us to measure the luminosity distance directly \citep{schutz1986}. Therefore, we can obtain the three-dimensional distribution of gravitational wave sources in the luminosity-distance space. However, cosmic inhomogeneities effectively perturb luminosity distances measured by the waveform analysis in several ways. At around $z\sim2$, the gravitational lensing effect is dominant compared to the Doppler effect due to peculiar velocities of objects \citep{kocsis2006,begnoni2024a}. As a result, the relation between the luminosity distance $D$ inferred from the gravitational waveform and the average luminosity distance $\bar{D}$ in a homogeneous and isotropic FLRW universe is given by
\begin{align}
D=\bar{D} \mu^{-1 / 2}\label{eq:luminosity_dis},
\end{align}
where $\mu$ is the magnification factor due to the gravitational lensing effect. This magnification factor $\mu$ can be Taylor expanded up to second order in the lensing convergence $\kappa$ and the shear $\gamma$ as
\begin{align}
\mu & =\frac{1}{(1-\kappa)^2-\gamma^2}\simeq 1+2 \kappa+3 \kappa^2+\gamma^2\label{eq:mu_app}.
\end{align}
Thus, the luminosity distance affected by the lensing effect can be expressed by substituting Eq.~\eqref{eq:mu_app} into Eq.~\eqref{eq:luminosity_dis} as
\begin{align}
D\simeq\bar{D} /\sqrt{1+2 \kappa+3 \kappa^2+\gamma^2}\label{eq:LuminosityApp}.
\end{align}
We note that the expansion up to the second order is needed to properly account for the lensing dispersion effect in calculating angular clustering signals.

The convergence $\kappa$ is a function of the position $\boldsymbol{\theta}$ on the celestial sphere and the redshift $z$ and is calculated by the line-of-sight projection of the matter density. Specifically, it is given by
\begin{align}\kappa(\boldsymbol{\theta},\chi_s)
& =\frac{3 \Omega_{\rm{m}} H_0^2}{2} \int_0^{\chi_{\rm{s}}} d \chi \frac{\left(\chi_{\rm{s}}-\chi\right) \chi}{\chi_{\rm{s}}} \frac{\delta_{\rm{m}}(\chi,\boldsymbol{\theta})}{a},
\end{align}
where $H_0$ is the Hubble constant, $\chi$ is the comoving radial distance, $\chi_{\rm{s}}$ is the comoving radial distance to the source, $\delta_{\mathrm{m}}(\chi,\boldsymbol{\theta})$ is the matter density fluctuation field, and $a$ is the scale factor.

The luminosity distance has errors arising from the measurement error of the gravitational wave signal, as well as degeneracies with the gravitational wave source mass, the binary orbital inclination, and the peculiar velocity. This error is estimated to be $\sigma_{\ln\rm{D}}=0.08$ for the third-generation gravitational wave detector Einstein Telescope, and $\sigma_{\ln\rm{D}}=0.02$ for DECIGO at $z\sim2$ \citep{camera2013}. Here, we assume that the luminosity distance $D_{\rm{obs}}$ obtained from the analysis of the observed waveform follows a log-normal distribution with the median $D$. Specifically, the probability distribution of $D_{\mathrm{obs}}$ is given by
\begin{align}
p\left(D_{\text {obs }} \mid D\right)=\frac{1}{\sqrt{2 \pi} \sigma_{\ln D}D_{\text {obs }}}\exp \left[-x^2\left(D_{\text {obs}}\right)\right]\label{eq:lognomal},
\end{align}
\begin{align}
x\left(D_{\mathrm{obs}}\right) \equiv \frac{\ln D_{\mathrm{obs}}-\ln D}{\sqrt{2} \sigma_{\ln D}}\label{eq:x_D_obs}.
\end{align}
The choice of the log-normal distribution is reasonable because errors of luminosity distances are given as relative errors. Gravitational wave observations detect gravitational wave signals with some signal-to-noise ratio cut, where the signal-to-noise represents the relative error of the signal. As a result, the error of the luminosity distance is expected to be a relative error and roughly follows the log-normal distribution.

From Eq.~\eqref{eq:LuminosityApp}, $x(D_{\rm{obs}})$ defined in Eq. \eqref{eq:x_D_obs} can be approximated as
\begin{align}
x(D_{\rm{obs}})&\simeq \bar{x}+\frac{1}{\sqrt{2} \sigma_{\ln D}}\left(\kappa+\frac{1}{2} \kappa^2+\frac{1}{2} \gamma^2\right)\label{eq:xDobs},
\end{align}
where
\begin{align}
\bar{x}=\frac{1}{\sqrt{2}\sigma_{\ln D}}\ln(D_{\rm{obs}}/\bar{D}).
\end{align}

\subsection{Projection of the Three-Dimensional Number Density Field onto the Two-Dimensional Celestial Sphere}\label{subsec:m_projection 3 to 2}
\subsubsection{Case$\;\rm{I}$ : Modeling Lensing Effects via the Taylor Expansion}\label{subsubsec:m_projection of GWs}
Using the luminosity distance described above, we construct the number density field of gravitational wave sources on the celestial sphere. Firstly, we assume a luminosity distance bin in the range $D_{\rm{min}}<D_{\rm{obs}}<D_{\rm{max}}$ and project positions of all gravitational wave sources with observed luminosity distances in this range onto the celestial sphere. The angular number density field $n^{\rm{w}}(\boldsymbol{\theta})$ of this sample is given by
\begin{align}
n^{\mathrm{w}}(\boldsymbol{\theta})=\int_0^{\infty} d z \frac{\chi^2}{H(z)} S(z) n_{\mathrm{GW}}(\boldsymbol{\theta}, z)\label{eq:ang_num_density},
\end{align}
where $H(z)$ is the Hubble parameter, $n_{\mathrm{GW}}(\boldsymbol{\theta}, z)$ is the three-dimensional number density field of gravitational wave sources and $S(z)$ is the selection function given by
\begin{align}
\nonumber S(z)&=\int_0^{\infty}dD_{\text {obs}} \Theta\left(D_{\text {obs}}-D_{\text {min}}\right) \Theta\left(D_{\text{max}}-D_{\text {obs}}\right)\\\nonumber&\quad\quad\quad\quad\times p\left(D_{\text {obs}} \mid D\right) \\
& =\frac{1}{2}\left(\operatorname{erfc}\left\{x\left(D_{\min }\right)\right\}-\operatorname{erfc}\left\{x\left(D_{\max }\right)\right\}\right),\end{align}
where we use the Heaviside step functions $\Theta\left(D_{\text {obs }}-D_{\min }\right)$ and $\Theta\left(D_{\max }-D_{\text {obs }}\right)$ to model the luminosity distance bin and $\operatorname{erfc}(z)$ is the complementary error function.

Secondly, by treating $\left(\kappa+ \kappa^2/2+\gamma^2/2\right)/\sqrt{2} \sigma_{\ln D}$ in Eq.~\eqref{eq:xDobs} as sufficiently small and expanding the selection function $S(z)$, we derive an expression for the angular number density field considering both observational errors and gravitational lensing effects. The Taylor expansion of the selection function reduces to
\begin{align}
\nonumber S(z)&=\frac{1}{2}\left(\operatorname{erfc}\left\{x\left(D_{\min }\right)\right\}-\operatorname{erfc}\left\{x\left(D_{\max}\right)\right\}\right)\\
&\simeq\bar{S}(z)+T(z)\left(\kappa+\frac{\kappa^2}{2}+\frac{\gamma^2}{2}\right)+U(z)\kappa^2,
\end{align}
where we assume $\kappa$ and $\gamma$ are sufficiently small and ignore the third and higher order terms, and $\bar{S}(z)$, $T(z)$, and $U(z)$ are defined as 
\begin{align}
\bar{S}(z) \equiv\; & \frac{1}{2} \left( 
\operatorname{erfc}\left\{\bar{x}(D_{\text{min}})\right\} 
- \operatorname{erfc}\left\{\bar{x}(D_{\text{max}})\right\} \right), \\
T(z) \equiv\; & \frac{1}{\sqrt{2\pi}\, \sigma_{\ln D}} \left[
    -\exp\left\{-\bar{x}^2(D_{\text{min}})\right\} \right. \nonumber \\
    & \left. \quad\quad\quad\quad\quad\quad\quad\quad\quad+ \exp\left\{-\bar{x}^2(D_{\text{max}})\right\}
\right], \\
U(z) \equiv\; & \frac{1}{2\sqrt{\pi}\, \sigma_{\ln D}^2} \left[
    \bar{x}(D_{\text{min}}) \exp\left\{-\bar{x}^2(D_{\text{min}})\right\} \right. \nonumber \\
    & \left. \quad\quad\quad\quad\quad\;- \bar{x}(D_{\text{max}}) \exp\left\{-\bar{x}^2(D_{\text{max}})\right\}
\right].
\end{align}
\begin{figure}[t]
\centering
\includegraphics[width=0.48\textwidth]{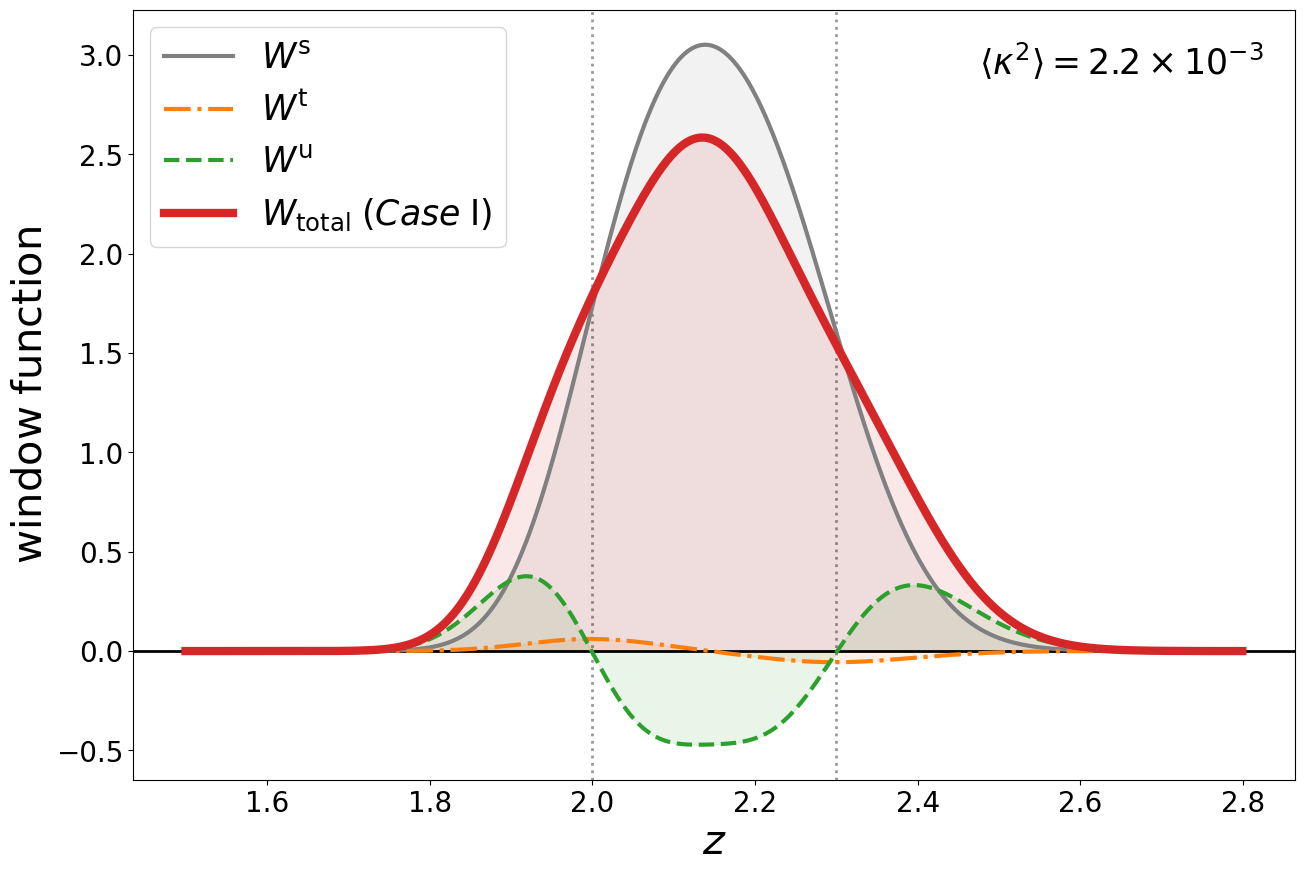}

\includegraphics[width=0.48\textwidth]{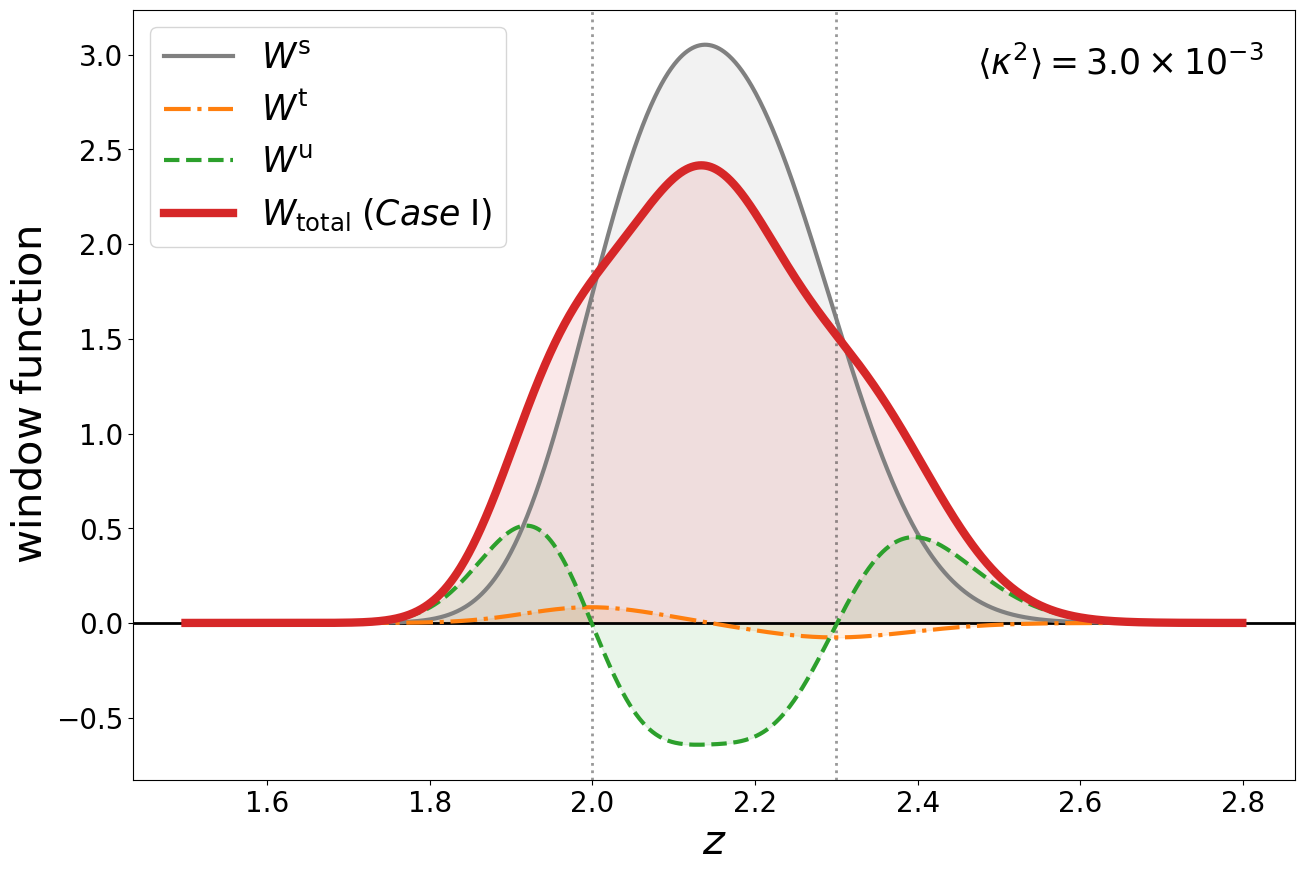}
\caption{\label{fig:epsart}
Effect of the lensing dispersion on the selection functions for constructing the angular density field. Results for $\langle\kappa^2\rangle=2.2\times10^{-3}$ (\textit{top}), and $\langle\kappa^2\rangle=3.0\times10^{-3}$ (\textit{bottom}) are shown. Dotted vertical lines indicate the redshift range corresponding to the observed luminosity distance bin, which we determine to satisfy $D_{\mathrm{min}}=\bar{D}(z=2.0)$ and $D_{\mathrm{max}}=\bar{D}(z=2.3)$. Gray thin solid lines show $W^{\rm{s}}$, which represents the spread of the selection function due to the observational error of gravitational wave sources $\sigma_{\ln D}=0.05$, which is assumed to follow a log-normal distribution. Due to gravitational lensing effects, the distribution changes, as shown by red thick solid lines. This change can be classified into two components, $W^{\rm{t}}$ and $W^{\rm{u}}$. Orange dash-dotted lines show $W^{\rm{t}}$, which shifts the overall distribution to a lower redshift than the original distribution. Green dashed lines show $W^{\rm{u}}$, which decreases the number density near the bin center and increases in regions away from the center, showing an effect of increasing the dispersion.}
\label{fig:distribution}
\end{figure}
Finally, we rewrite the angular number density field $n^{\mathrm{w}}(\boldsymbol{\theta})$ in terms of the selection functions $\bar{S}(z),T(z),U(z)$, and the three-dimensional number density field of gravitational wave sources $n_{\rm{GW}}(\boldsymbol{\theta},z)$ as
\begin{align}
n^{\mathrm{w}}(\boldsymbol{\theta})&=\nonumber\int_0^{\infty} d z \frac{\chi^2}{H(z)}\left[{\bar{S}}(z)\right.\\&\quad\quad\left.+T(z)\left(\kappa+\frac{\kappa^2}{2}+\frac{\gamma^2}{2}\right)+U(z)\kappa^2\right]{n}_{\rm{GW}}(\boldsymbol{\theta},z)\label{eq:ang_num_density2}.
\end{align}
From this expression, we can define the two-dimensional density fluctuation $\delta_{\rm{GW}}(\boldsymbol{\theta}, z)$ needed for the auto- and cross-correlation analysis. Specifically, it is defined as
\begin{align}
\delta_{\rm{GW}}(\boldsymbol{\theta}) \equiv \frac{n^{\rm{w}}(\boldsymbol{\theta})-\bar{n}^{\rm{w}}}{\bar{n}^{\rm{w}}}.
\end{align}
 The average number density of gravitational wave sources $\bar{n}^{\mathrm{w}}$ in the above equation can be obtained by taking the average of Eq. \eqref{eq:ang_num_density2} as
\begin{align}
\nonumber\bar{n}^{\mathrm{w}}&=\left\langle\int_0^{\infty} d z \frac{\chi^2}{H(z)} S(z) n_{\mathrm{GW}}(\boldsymbol{\theta},z)\right\rangle\\
&=\int_0^{\infty} dz \frac{\chi^2}{H(z)}\left[{\bar{S}}(z)-T(z)\langle\kappa^2\rangle+U(z)\langle\kappa^2\rangle\right]\bar{n}_{\rm{GW}}(z),\label{eq:average number density}\end{align}
 where $\bar{n}_{\mathrm{GW}}(z)$ is the three-dimensional average number density of gravitational wave sources at redshift $z$. When taking the average, we use the properties of the lensing shear and convergence $\langle\kappa\rangle=-2\langle\kappa^2\rangle$ and $\langle\gamma^2\rangle=\langle\kappa^2\rangle$, based on the results of ray-tracing simulations \citep{takahashi2011}. The two-dimensional density fluctuation of gravitational wave sources on the celestial sphere $\delta^{\rm{2D},w}(\boldsymbol{\theta})$ is calculated as
\begin{align}
\nonumber\delta^{\rm{2D},w}(\boldsymbol{\theta})
\nonumber&=\int_0^{\infty}dz\left[W^{\rm{s}}(z)\delta_{\rm{GW}}(\boldsymbol{\theta},z)\right.\\
\nonumber&\quad\quad\left.+W^{\rm{t}}(z)\left(\kappa+\frac{\kappa^2}{2}+\frac{\gamma^2}{2}+\langle\kappa^2\rangle\right)\right.\\
\nonumber&\quad\quad\left.+W^{\rm{t}}(z)\delta_{\rm{GW}}(\boldsymbol{\theta},z)\left(\kappa+\frac{\kappa^2}{2}+\frac{\gamma^2}{2}\right)\right.\\
&\quad\quad\left.+W^{\rm{u}}(z)\delta_{\rm{GW}}(\boldsymbol{\theta},z)\kappa^2+W^{\rm{u}}(z)(\kappa^2-\langle\kappa^2\rangle)\right],\label{eq:delta_2D_w}\end{align}
where
\begin{align}
& W^{\rm{s}}(z) \equiv \frac{1}{\bar{n}^{\rm{w}}} \frac{\chi^2}{H(z)} \bar{n}_{\rm{GW}}(z) \bar{S}(z), \\
& W^{\rm{t}}(z) \equiv \frac{1}{\bar{n}^{\rm{w}}} \frac{\chi^2}{H(z)} \bar{n}_{\rm{GW}}(z) T(z) ,\\
& W^{\rm{u}}(z) \equiv \frac{1}{\bar{n}^{\rm{w}}} \frac{\chi^2}{H(z)} \bar{n}_{\rm{GW}}(z) U(z).
\end{align}
As shown in Fig.~\ref{fig:distribution}, the first term $W^{\rm{s}}$ in Eq.~\eqref{eq:delta_2D_w} represents the spatial inhomogeneity of gravitational wave sources, while the subsequent terms $W^{\rm{t}},W^{\rm{u}}$ represent changes in the apparent distribution of gravitational wave sources due to gravitational lensing effects. 

\subsubsection{Case$\;\rm{I\hspace{-1.2pt}I}$ : Modeling Lensing Effects via the Log-normal Dispersion}\label{subsubsec:m_case2}
\begin{figure}[t]
\centering
\includegraphics[width=0.48\textwidth]{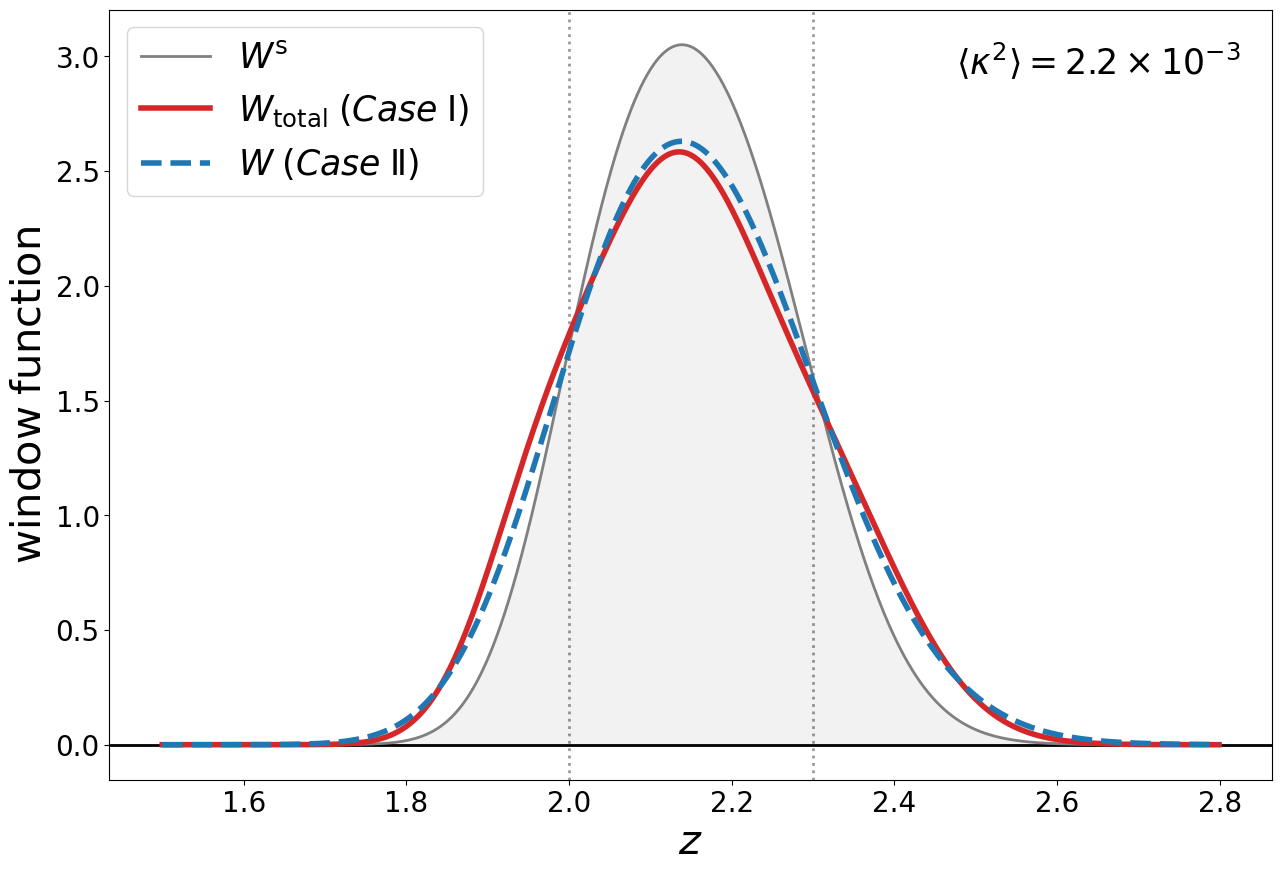}
\includegraphics[width=0.48\textwidth]{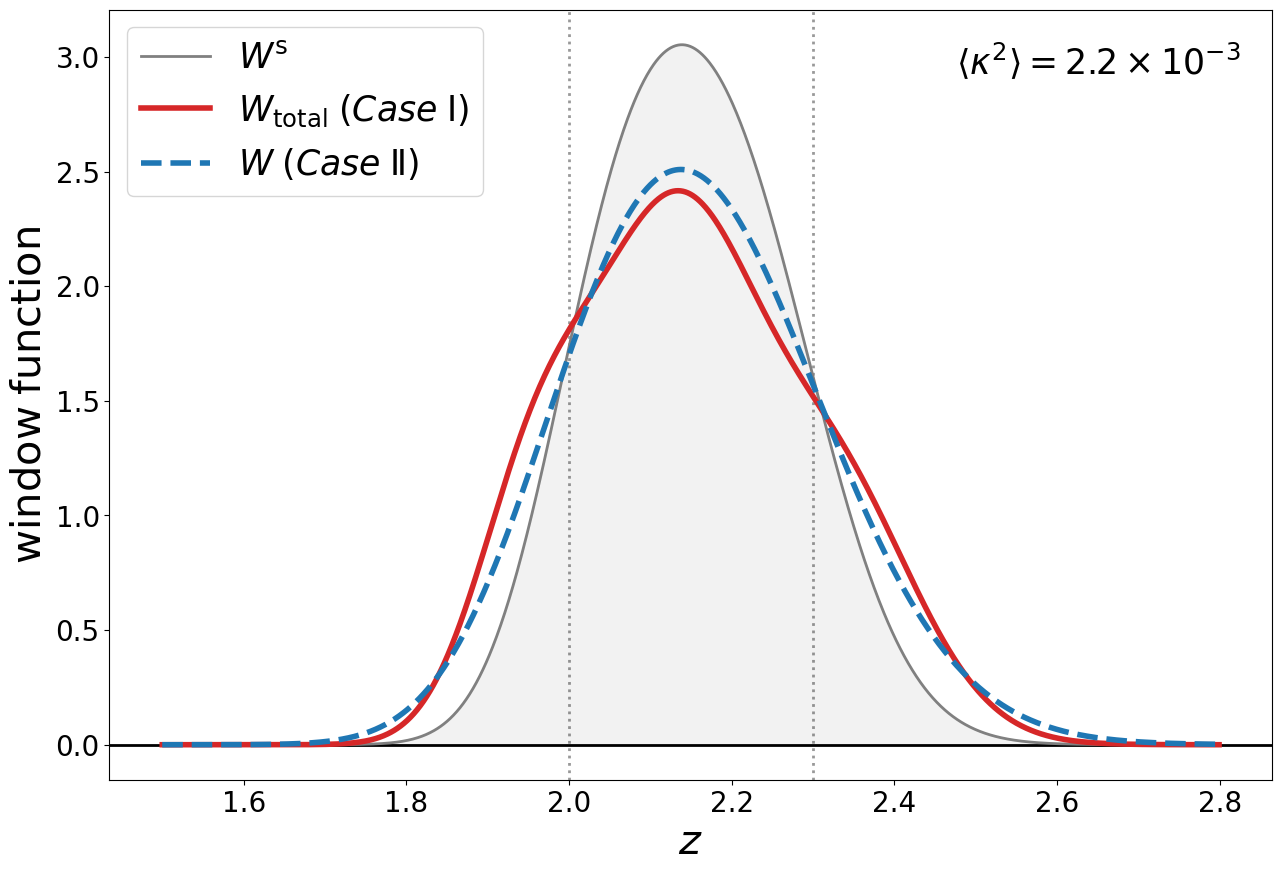}
\caption{\label{fig:epsart}
Differences in the selection functions between Case$\;\rm{I}$ and Case$\;\rm{I\hspace{-1.2pt}I}$. Selection functions for Case I  and Case$\;\rm{I\hspace{-1.2pt}I}$ are shown by red solid lines and blue dashed lines, respectively. The top panel shows the result for $\langle\kappa^2\rangle=2.2\times10^{-3}$ , and the bottom panel is for $\langle\kappa^2\rangle=3.0\times10^{-3}$. Dotted vertical lines indicate the range of the observed luminosity bin assumed for the calculation (see also the caption of Fig.~\ref{fig:distribution}). }
\label{fig:distribution_1_2}
\end{figure}
So far, we resort to the Taylor expansion to include the effect of the lensing dispersion. There is another approach to approximately account for gravitational lensing effects by adding the dispersion of the lensing convergence to the standard deviation of the log-normal distribution (e.g.,~\citep{adamek2019}). Specifically,
\begin{align}
p_{\rm{I\hspace{-1.2pt}I}}\left(D_{\mathrm{obs}} \mid D\right)=\frac{1}{\sqrt{2 \pi} \sigma_{\mathrm{tot}}} \exp \left[-x_{\rm{I\hspace{-1.2pt}I}}^2\left(D_{\mathrm{obs}}\right)\right],
\end{align}
where
\begin{align}
x_{\rm{I\hspace{-1.2pt}I}}\left(D_{\mathrm{obs}}\right) &\equiv \frac{\ln D_{\mathrm{obs}}-\ln D}{\sqrt{2} \sigma_{\rm{tot}}},\\
\sigma^2_{\rm{tot}}&\equiv\sigma_{\ln D}^2+\sigma^2_{\langle\kappa^2\rangle},
\end{align}
and $\sigma_{\langle\kappa^2\rangle}^2=\langle\kappa^2\rangle$. Hereafter, we refer to the method using the Taylor expansion to include lensing effects on luminosity distances as Case$\;\rm{I}$, and the method adding the lensing dispersion to the standard deviation of the log-normal distribution as Case$\;\rm{I\hspace{-1.2pt}I}$. Fig.~\ref{fig:distribution_1_2} shows the differences between these two methods. 

A significant advantage of Case$\;\rm{I}$ is that we can incorporate the relation between the average and the dispersion of the lensing convergence, $\langle\kappa\rangle=-2\left\langle\kappa^2\right\rangle$, and separate gravitational lensing effects into the effect of shifting the selection function and the effect of increasing its variance, as shown in Fig.~\ref{fig:distribution}. This approach enables us to understand the lensing effects of the selection function more clearly. 

However, when deriving the angular power spectrum from the fluctuations in the number density field of gravitational wave sources following Eq.~\eqref{eq:delta_2D_w}, multiple contributing terms arise, each of which must be carefully evaluated to assess its relative impact on the angular power spectrum. On the other hand, Case$\;\rm{I\hspace{-1.2pt}I}$ is simple and is easy to implement, whereas it ignores some effects such as the shift of the selection function seen in Case$\;\rm{I}$. The difference between these two approaches becomes more significant with increasing the value of the lensing dispersion. Notably, the effect of $W^{\rm{u}}$ in Case$\;\rm{I}$ becomes more significant for the larger lensing dispersion, leading to larger differences between Case$\;\rm{I}$ and Case$\;\rm{I\hspace{-1.2pt}I}$. They show minimal differences for the smaller lensing dispersion. Therefore, understanding the differences between these two methods, both of which involve approximations, is important for the future analysis of the effect of the lensing dispersion on the angular clustering of gravitational wave sources. Consequently, in this paper we present results for both Case$\;\rm{I}$ and Case$\;\rm{I\hspace{-1.2pt}I}$ and compare these two approaches.

\subsection{Projection of Spectroscopic Galaxies}\label{subsec:m_projection of spectroscopic galaxies}
For the cross-correlation analysis, we consider a galaxy sample with spectroscopic redshift measurements. Selecting the galaxies with $z_{\min}<z<z_{\max }$, the angular number density field of spectroscopic galaxies is obtained as
\begin{align}
n^{\mathrm{g}}(\boldsymbol{\theta})=\int_0^{\infty} d z \frac{\chi^2}{H(z)} \Theta\left(z-z_{\min}\right) \Theta\left(z_{\max}-z\right) n_{\mathrm{g}}(\boldsymbol{\theta}, z)\label{eq:ang_num_density_g}.
\end{align}
From this expression, we can define the two-dimensional density fluctuation $\delta_{\rm{g}}(\boldsymbol{\theta}, z)$ . Specifically, it is defined as
\begin{align}
\delta_{\rm{g}}(\boldsymbol{\theta}) \equiv \frac{n^{\rm{g}}(\boldsymbol{\theta})-\bar{n}^{\rm{g}}}{\bar{n}^{\rm{g}}}.
\end{align}
 The average number density of spectroscopic galaxies $\bar{n}^{\mathrm{g}}$ can be obtained by taking the average of Eq. \eqref{eq:ang_num_density_g} as
\begin{align}\bar{n}^{\mathrm{g}}&=\int_0^{\infty} d z \frac{\chi^2}{H(z)} \bar{n}_{\mathrm{g}}(z) \Theta\left(z-z_{\min}\right) \Theta\left(z_{\max}-z\right),\label{eq:galaxy_ang_density}
\end{align}
where $\bar{n}_{\rm{g}}(z)$ is the average density of spectroscopic galaxies at redshift $z$, and the selection function is constructed based on the Heaviside function. The two-dimensional galaxy density fluctuation $\delta^{2 \mathrm{D}, \mathrm{g}}(\boldsymbol{\theta})$ is calculated as
\begin{align}
\delta^{2 \mathrm{D}, \mathrm{g}}(\boldsymbol{\theta})=\int_0^{\infty} d z W^{\mathrm{g}}(z) \delta_{\mathrm{g}}(\boldsymbol{\theta}, z),\label{eq:galaxy_fluc}
\end{align}
where
\begin{align}
W^{\rm{g}}(z) \equiv \frac{1}{\bar{n}^{\mathrm{g}}} \frac{\chi^2}{H(z)} \bar{n}_{\mathrm{g}}(z) \Theta\left(z-z_{\min}\right) \Theta\left(z_{\max}-z\right).
\end{align}

\subsection{Linear bias of Gravitational Wave Sources}\label{subsec:m_linear bias}
Throughout this paper, we assume linear bias that are a good approximation for binary black holes as long as we focus on clustering at large scales (e.g., \citep{namikawa2016b}). While the origin of binary black holes is not yet known, it is reasonable to assume that they more or less follow the galaxy distribution on large scales. The linear bias is a valid assumption in the sense that compact objects such as binary black holes correlate with the galaxy distribution and trace the large-scale structure of the Universe on large scales. 
        
\subsection{Auto-correlation Angular Power Spectrum}\label{subsec:m_angular power spectra}
This section presents the auto-correlation angular power spectrum of binary black holes. We provide formulae for both Case$\;\rm{I}$ and Case$\;\rm{I\hspace{-1.2pt}I}$.
\subsubsection{\texorpdfstring{Case$\;\rm{I}$}{Case I}}\label{subsubsec:Case1}
The derivation of the formula for Case I is based on several key assumptions and approximations. First, we assume that the number density fluctuation of gravitational wave sources denoted as \(\delta_{\rm{GW}}(z)\) and the lensing convergence \(\kappa\) are independent. This leads to the relationship \(\langle\delta_{\rm{GW}}\kappa\rangle = \langle\delta_{\rm{GW}}\rangle\langle\kappa\rangle\). This assumption is justified because the source number density fluctuation is defined within a specific luminosity distance bin, while the lensing convergence \(\kappa\) accumulates density fluctuations along the entire line of sight from the source to the observer. Second, we ignore the terms that involve third- or higher-order products of $\kappa$ and $\gamma$, treating them as higher-order small quantities. Third, we also ignore the terms of the form 
$W_j^{\rm{t}}\left(z^{\prime}\right)\left\langle\delta_{\rm{GW}}(\boldsymbol\theta,z) \delta_{\rm{GW}}\left(\boldsymbol\theta^{\prime},z^{\prime}\right)\right\rangle\left\langle \kappa(\boldsymbol\theta,z) \kappa\left(\boldsymbol\theta^{\prime},z^{\prime}\right)\right\rangle$ since these are expected to be significantly smaller than the other terms. This expectation arises because our analysis primarily focuses on relatively large angular scales, specifically in the range $\ell \approx 100$ to $\ell \approx 1000$. On these scales, the auto-correlation functions of the lensing convergence and the gravitational wave source number density can be approximated as
\begin{align}
&\left\langle \kappa(\boldsymbol\theta,z) \kappa\left(\boldsymbol\theta^{\prime},z^{\prime}\right)\right\rangle\simeq \ell^2C^{\kappa\kappa}(\ell)\left.\right|_{\ell\simeq100}\simeq10^{-5},\\
&\langle\delta_{\rm{GW}}(\boldsymbol\theta,z)\delta_{\rm{GW}}\left(\boldsymbol\theta^{\prime},z^{\prime}\right)\rangle\simeq \ell^2C^{\mathrm{ww}}(\ell)\left.\right|_{\ell\simeq100}\simeq10^{-2},
\end{align}
leading to the product
\begin{align}
\left\langle\delta_{\rm{GW}}(\boldsymbol\theta,z) \delta_{\rm{GW}}\left(\boldsymbol\theta^{\prime},z^{\prime}\right)\right\rangle\left\langle \kappa(\boldsymbol\theta,z) \kappa\left(\boldsymbol\theta^{\prime},z^{\prime}\right)\right\rangle\simeq10^{-7},
\end{align}
which is much smaller than the typical lensing variance $\langle\kappa^2\rangle\simeq10^{-3}-10^{-2}$. Finally, we adopt the Limber approximation in the derivation. Under these assumptions, we obtain the auto-correlation angular power spectrum of gravitational wave sources as

\begin{align}
C^{\mathrm{w} \mathrm{w}}(\ell)=C^{\mathrm{s}\mathrm{s}}(\ell)+C^{\mathrm{s} \mathrm{t}}(\ell)+C^{\mathrm{s} \mathrm{u}}(\ell)+C^{\mathrm{t}\mathrm{t}}(\ell),\label{eq:Cww_Case1}
\end{align}
\begin{align}
C^{\rm{s} \rm{s}}(\ell)=\int_0^{\infty} d z W^{\rm{s}}(z) W^{\rm{s}}(z) \frac{H(z)}{\chi^2} b_{\rm{GW}}^2P_{\rm{m}}\left(\frac{\ell+1 / 2}{\chi}, z\right),
\end{align}
\begin{align}
& \nonumber C^{\rm{s} \rm{t}}(\ell)=-2 \int_0^{\infty} d z W^{\rm{s}}(z) W^{\rm{t}}(z)\\&\quad\quad\quad\quad\quad\times \frac{H(z)}{\chi^2} b_{\rm{GW}}^2 P_{\rm{m}}\left(\frac{\ell+1/2}{\chi}, z\right) \langle {\kappa(z, \boldsymbol\theta)}^{2}\rangle\label{eq:Cst},
\end{align}
\begin{align}
\nonumber C^{\rm{s}\rm{u}}(\ell)&=2\int^{\infty}_0dz\: W^{\rm{s}}(z) W^{\rm{u}}(z) \\&\quad\quad\quad\quad\times\frac{H(z)}{\chi^2} b_{\rm{GW}}^2 P_{\rm{m}}\left(\frac{\ell+1 / 2}{\chi}, z\right) \langle {\kappa(z, \boldsymbol\theta)}^{2}\rangle,
\end{align}
\begin{align}
\nonumber C&^{\rm{t}\rm{t}}(\ell)=\int^{\infty}_0dz\left[W^{\rm{t}}(z) W^{\rm{t}}(z) \frac{H(z)}{\chi^2} b_{\rm{GW}}^2 P_{\rm{m}}\left(\frac{\ell+1 / 2}{\chi}, z\right)\right.\\\nonumber\times&\left.\int_0^z d z^{\prime \prime} W^{\kappa}\left(z^{\prime \prime} ; z\right) W^{\kappa}(z^{\prime \prime} ; z)\frac{H(z^{\prime \prime})}{\chi^{\prime \prime 2}} P_{\rm{m}}\left(\frac{\ell+1/2}{\chi^{\prime \prime}}, z^{\prime \prime}\right)\right]\\\nonumber+& \int_0^{\infty} d z \int_0^{\infty} d z^{\prime}\:W^{\rm{t}}(z) W^{\rm{t}}(z^\prime)\\\nonumber\times&\int_0^{\min \left(z, z^{\prime}\right)} d z^{\prime \prime} W^{\kappa}\left(z^{\prime \prime} ; z\right) W^{\kappa}(z^{\prime \prime} ; z^{\prime})\\\times&\frac{H(z^{\prime \prime})}{\chi^{\prime \prime 2}} P_{\rm{m}}\left(\frac{\ell+1/2}{\chi^{\prime \prime}}, z^{\prime \prime}\right),
\end{align}
where
\begin{align}
W^\kappa(z;z_s)\equiv\frac{3 \Omega_{\mathrm{m} } H_0^2}{2} \frac{\left(\chi_{\mathrm{s}}-\chi\right) \chi}{\chi_{\mathrm{s}}}(1+z).
\end{align}
The first term $C^{\mathrm{s} \mathrm{s}}$ originates from the intrinsic clustering of binary black holes. In contrast, the second term $C^{\mathrm{s} \mathrm{t}}$ and the third term $C^{\mathrm{s}\mathrm{u}}$ arise from the correlation between the weak gravitational lensing effect on the luminosity distance of gravitational wave sources and the clustering of binary black holes. The fourth term arises from the weak lensing effect on the luminosity distance of gravitational wave sources.

Here, we discuss the matter power spectrum $P_{\mathrm{m}}(k)$. Our target scales are large scales, where the two-halo term dominates the auto-correlation angular power spectrum. Therefore, we apply the linear matter power spectrum for $C^{\mathrm{s}\mathrm{s}}(\ell)$, $C^{\mathrm{s} \mathrm{t}}(\ell)$, and $C^{\mathrm{s} \mathrm{u}}(\ell)$. The linear matter power spectrum is calculated using the transfer function derived in \citet{eisenstein1998}. On the contrary, for $C^{\mathrm{t}\mathrm{t}}(\ell)$, we apply the nonlinear matter power spectrum since $C^{\mathrm{t}\mathrm{t}}(\ell)$ includes density fluctuations at various scales between the binary black holes and the observer. The nonlinear matter power spectrum is calculated following \citet{takahashi2012}. 
\subsubsection{\texorpdfstring{Case$\;\rm{I\hspace{-1.2pt}I}$}{Case II}}\label{subsubsec:Case2}
Following the definition of Case$\;\rm{I\hspace{-1.2pt}I}$ in Sec.~\ref{subsubsec:m_case2}, we also derive the formula of Case$\;\rm{I\hspace{-1.2pt}I}$ as
\begin{align}
    C^{\rm{ww}}(\ell)=\int_0^{\infty} d z W^{\mathrm{s}}_{\rm{I\hspace{-1.2pt}I}}(z) W^{\mathrm{s}}_{\rm{I\hspace{-1.2pt}I}}(z) \frac{H(z)}{\chi^2} b_{\mathrm{GW}}^2 P_{\mathrm{m}}\left(\frac{\ell+1 / 2}{\chi}, z\right),
\end{align}
where
\begin{align}
W^{\mathrm{s}}_{\rm{I\hspace{-1.2pt}I}}(z) \equiv &\frac{1}{\bar{n}^{\mathrm{w}}} \frac{\chi^2}{H(z)} \bar{n}_{\mathrm{GW}}(z) S_{\rm{I\hspace{-1.2pt}I}}(z),\\
\nonumber S_{\rm{I\hspace{-1.2pt}I}}(z)= & \int_0^{\infty} d D_{\mathrm{obs}} \Theta\left(D_{\mathrm{obs}}-D_{\min }\right) \Theta\left(D_{\max }-D_{\mathrm{obs}}\right) \\
&\nonumber\quad\quad\quad\quad\times p_{\rm{I\hspace{-1.2pt}I}}\left(D_{\mathrm{obs}} \mid D\right)\\
= & \frac{1}{2}\left(\operatorname{erfc}\left\{x_{\rm{I\hspace{-1.2pt}I}}\left(D_{\min }\right)\right\}-\operatorname{erfc}\left\{x_{\rm{I\hspace{-1.2pt}I}}\left(D_{\max }\right)\right\}\right).
\end{align}
\subsubsection{Galaxy Auto-Correlation Angular Power Spectrum
}
Using Eq.~\eqref{eq:galaxy_fluc}, the galaxy auto-correlation angular power spectrum is derived similarly to the gravitational wave source auto-correlation angular power spectrum as
\begin{align}
C^{\mathrm{g}\mathrm{g}}(\ell)=\int_0^{\infty} d z\left[W^{\mathrm{g}}(z)\right]^2 \frac{H(z)}{\chi^2} b_{\mathrm{g}}^2 P_{\mathrm{m}}\left(\frac{\ell+1 / 2}{\chi} ; z\right),
\end{align}
where $b_{\rm{g}}$ is the linear bias of spectroscopic galaxies. In the following analysis, we always assume that the linear bias of spectroscopic galaxies is already known from observations of the galaxy auto-correlation angular power spectrum.

\subsection{Cross-correlation Angular Power Spectrum}\label{sec:cross-corelation}
In this section, we present the cross-correlation angular power spectrum between gravitational wave sources and spectroscopic galaxies for both Case$\;\rm{I}$ and Case$\;\rm{I\hspace{-1.2pt}I}$.
As shown in Sec.~\ref{subsec:m_angular power spectra}, the auto-correlation angular power spectrum depends on both the lensing dispersion and the linear bias of gravitational wave sources. To better constrain these parameters, we consider a joint analysis that combines the auto-correlation of gravitational wave sources with their cross-correlation with spectroscopic galaxies.
\subsubsection{\texorpdfstring{Case$\;\rm{I}$}{Case I}}\label{subsubsec:Case1}
For  Case$\;\rm{I}$, the cross-correlation angular power spectrum
 between binary black holes and spectroscopic galaxies is given by
\begin{align}
C^{\mathrm{w} \mathrm{g}}(\ell)=C^{\mathrm{s}\mathrm{g}}(\ell)+C^{\mathrm{t}\mathrm{g}}(\ell)+C^{\mathrm{u}\mathrm{g}}(\ell)\label{eq:Cwg_Case1},
\end{align}
\begin{align}
\nonumber C^{\mathrm{s} \mathrm{g}}(\ell)= & \int_0^{\infty} d z W^{\mathrm{s}}(z) W^{\mathrm{g}}(z)\\
&\quad\times\frac{H(z)}{\chi^2}  b_{\mathrm{g}}b_{\mathrm{GW}}P_{\mathrm{m}}\left(\frac{\ell+1 / 2}{\chi} ; z\right),
\end{align}
\begin{align}
\nonumber C^{\mathrm{t}\mathrm{g}}(\ell)=- & \int_0^{\infty} d z W^{\mathrm{t}}(z) \int_0^z d z^{\prime}W^{\mathrm{g}}\left(z^{\prime}\right) 
\\&\times \frac{H\left(z^{\prime}\right)}{\chi^{\prime 2}} b_{\mathrm{g}}b_{\mathrm{GW}} P_{\mathrm{m}}\left(\frac{\ell+1 / 2}{\chi} ; z^{\prime}\right) \langle {\kappa(z, \boldsymbol\theta)}^{2}\rangle\label{eq:Ctg},
\end{align}
\begin{align}
\nonumber C^{\mathrm{u}\mathrm{g}}(\ell)= & \int_0^{\infty} d z W^{\mathrm{u}}(z) \int_0^z d z^{\prime} W^{\mathrm{g}}\left(z^{\prime}\right)\\
&\quad\times \frac{H\left(z^{\prime}\right)}{\chi^{\prime 2}} b_{\mathrm{g}}b_{\mathrm{GW}} P_{\mathrm{m}}\left(\frac{\ell+1 / 2}{\chi} ; z^{\prime}\right) \langle {\kappa(z, \boldsymbol\theta)}^{2}\rangle.
\end{align}

Here, we use the linear matter power spectrum $P_{\mathrm{m}}(k ; z)$ for all the three components $C^{\mathrm{s}\mathrm{g}}$, $C^{\mathrm{t}\mathrm{g}}$, and $C^{\mathrm{u}\mathrm{g}}$. The first term $C^{\mathrm{s} \mathrm{g}}$ originates from the intrinsic clustering of binary black holes and galaxies. In contrast, the second term $C^{\mathrm{t} \mathrm{g}}$ and the third term $C^{\mathrm{u}\mathrm{g}}$ arise from the correlation between the weak lensing effect on the luminosity distance of gravitational wave sources and spectroscopic galaxies. Since the weak lensing depends on all matter density fluctuations integrated along the line of sight, it induces non-negligible cross-correlations even between widely separated luminosity distance bins and redshift bins. In particular, $C^{\mathrm{t} \mathrm{g}}$ has the potential to generate anti-correlations.
\subsubsection{\texorpdfstring{Case$\;\rm{I\hspace{-1.2pt}I}$}{Case II}}\label{subsubsec:Case2}
For Case$\;\rm{I\hspace{-1.2pt}I}$, the cross-correlation angular power spectrum
 is given by
\begin{align}
\nonumber C^{\mathrm{w} \mathrm{g}}(\ell)= & \int_0^{\infty} d z W_{\rm{I\hspace{-1.2pt}I}}^{\mathrm{s}}(z) W^{\mathrm{g}}(z)\\&\quad\times\frac{H(z)}{\chi^2}  b_{\mathrm{g}}b_{\mathrm{GW}}P_{\mathrm{m}}\left(\frac{\ell+1 / 2}{\chi} ; z\right).
\end{align}
Finally, we note that the formalism presented so far is not limited to binary black holes, but can also be applied to e.g., binary neutron stars by appropriately modifying relevant parameters such as the linear bias and the number density.
    
\subsection{Analysis Methods}\label{subsec:analysis methods}
We present the expression of the signal-to-noise ratios of the auto- and cross-correlation angular power spectra that are used to examine whether the signals are observable. We then introduce a Fisher matrix analysis that is used to quantify how the combination of these angular power spectra helps better constrain the linear bias and the lensing dispersion.
 
\subsubsection{Signal-to-Noise Ratio}\label{subsubsec:snr}
The signals in this study are the auto-correlation angular power spectrum $C^{\rm{ww}}(\ell)$ and the cross-correlation angular power spectrum $C^{\rm{wg}}(\ell)$. The covariance matrix describes uncertainties of the measured angular power spectra. This incorporates contributions from both the cosmic variance and the shot noise, and under the Gaussian approximation, it is given by

\begin{align}
&\nonumber\operatorname{Cov}\left[C^{i j}(\ell), C^{m n}\left(\ell^{\prime}\right)\right]\\&\quad\quad\quad\quad= \frac{4 \pi}{\Omega_{\mathrm{s}}} \frac{\delta_{\ell \ell^{\prime}}}{(2 \ell+1) \Delta \ell}
\left(\tilde{C}^{i m} \tilde{C}^{j n}+\tilde{C}^{i n} \tilde{C}^{j m}\right).
\end{align}
Here, the indices $i,j,\cdots$ run over $\rm{w}$ and $\rm{g}$ to denote the type of the angular power spectrum, $\Omega_{\rm{s}}$ is the survey area, $\Delta \ell$ is the width of the $\ell$ bin, and $\tilde{C}$ represents the power spectrum including the shot noise as
\begin{align}
\tilde{C}^{i j}=C^{i j}+\delta_{i j} \frac{1}{\bar{n}^i},
\end{align}
where $\bar{n}^i$ represents the number densities of gravitational wave sources and spectroscopic galaxies given by equations Eq.~\eqref{eq:average number density} and Eq.~\eqref{eq:galaxy_ang_density}, respectively.

The components of the covariance matrix are calculated as follows. Firstly, the covariance of the auto-correlation of the angular power spectrum $C^{\rm{ww}}(\ell)$ for gravitational wave sources is given by
\begin{align}
\operatorname{Cov}\left[C^{\rm{ww}}(\ell), C^{\rm{ww}}(\ell^{\prime})\right]=\frac{4 \pi}{\Omega_{\mathrm{s}}} \frac{2\delta_{\ell \ell^{\prime}}}{(2 \ell+1) \Delta \ell}\left(C^{\rm{ww}}+\frac{1}{\bar{n}^{\rm{w}}}\right)^2.
\end{align}
Next, the covariance of the cross-correlation angular power spectrum $C^{\rm{wg}}(\ell)$ between gravitational wave sources and spectroscopic galaxies is given by

\begin{align}
&\nonumber\operatorname{Cov}\left[C^{ \rm{w}\rm{g}}(\ell), C^{\rm{w} \rm{g}}(\ell^{\prime})\right]=\frac{4 \pi}{\Omega_{\mathrm{s}}}\frac{\delta_{\ell \ell^{\prime}}}{(2 \ell+1) \Delta \ell}\\
&\quad\quad\quad\quad\times\left[\left(C^{\rm{g} \rm{g}}+\frac{1}{\bar{n}^{\rm{g}}}\right)\left(C^{\rm{w} \rm{w}}+\frac{1}{\bar{n}^{\rm{w}}}\right)+\left(C^{ \rm{w}\rm{g}}\right)^2\right].
\end{align}
Finally, the cross-covariance between the auto- and cross-correlation angular power spectra, which is relevant for the joint analysis, is given by
\begin{align}
\nonumber\operatorname{Cov}&[C^{\mathrm{ww}}(\ell), C^{\mathrm{wg}}(\ell^{\prime})]\\
&\quad=\frac{4 \pi}{\Omega_{\mathrm{s}}} \frac{2\delta_{\ell \ell^{\prime}}}{(2 \ell+1) \Delta \ell}\left(C^{\mathrm{ww}}+\frac{1}{\bar{n}^{\mathrm{w}}}\right) C^{\mathrm{wg}}\label{eq:cross-covariance}.
\end{align}

Using this covariance matrix, the signal-to-noise ratio is expressed as \begin{align}S/N=\sqrt{\sum^{\ell_{\rm{max}}}_{\ell}{C^{ij}({\ell})}\left[\operatorname{Cov}\left[C^{i j}(\ell), C^{m n}(\ell)\right]\right]^{-1} {C^{mn}({\ell})}}.
\end{align}
The behavior of the signal-to-noise ratio varies depending on the signal strength and the shot noise. For example, when the signal $C^{\rm{ww}}({\ell})$ is sufficiently large compared to the shot noise, i.e., when $C^{\mathrm{ww}}({\ell})\gg1/\bar{n}^{\mathrm{w}}$, the signal-to-noise ratio of the auto-correlation angular power spectrum of gravitational wave sources is
\begin{align}
(S/N)_{\rm{ww}}=\nonumber&\sqrt{\sum^{\ell_{\rm{max}}}_{\ell}\frac{{C^{\mathrm{ww}}}^2}{\frac{2}{2\ell+1}\left({{C^{\mathrm{ww}}}}+{1}/{\bar{n}^{\rm{w}}}\right)^2}}\\\simeq&\sqrt{\sum_{\ell}^{\ell_{\rm{max}}}\ell}\simeq\ell_{\rm{max}},
\end{align}
which increases monotonically with increasing $\ell_{\rm{max}}$. On the other hand, in the case when $C^{\rm{ww}}({\ell})$ is sufficiently small compared to the shot noise, i.e., when $C^{\rm{ww}}({\ell})\ll1/\bar{n}^{\rm{w}}$,
\begin{align}(S/N)_{\rm{ww}}=&\nonumber\sqrt{\sum^{\ell_{\rm\max}}_\ell\frac{{{C^{\mathrm{ww}}}^2}}{\frac{2}{2\ell+1}\left({{C^{\mathrm{ww}}}}+{1}/{\bar{n}^{\rm{w}}}\right)^2}}\\\simeq&\sqrt{\sum_{\ell}^{\ell_{\rm{max}}} \ell {\bar{n}^{\mathrm{w}}{}}^2{C^{\mathrm{ww}}}^2}.\label{eq:snr_approx}\end{align}
In this case, the behavior of the signal-to-noise ratio is influenced by the dependence of $C^{\rm{ww}}(\ell)$ on  $\ell$.  It is also sensitive to the average number density of gravitational wave sources, which enters the calculation of the signal-to-noise ratio through the shot noise contribution.
\subsubsection{Fisher Analysis}\label{subsubsec:fisher analysis}
The Fisher information matrix is expressed as
\begin{align}
F_{\alpha \beta}=\sum_{\ell} \sum_{i, j, m, n} \frac{\partial C^{i j}}{\partial p_\alpha}\left[\operatorname{Cov}\left(C^{i j}, C^{m n}\right)\right]^{-1} \frac{\partial C^{m n}}{\partial p_\beta},
\end{align}
where the indices $i, j, \cdots$ run over w and g to denote the types of the angular power spectrum, and $p_\alpha$ represents parameters. The $1\sigma$ error for each parameter is obtained by
\begin{align}
\sigma\left(p_\alpha\right)=\sqrt{\left(F^{-1}\right)_{\alpha \alpha}}.
\end{align}
The parameters in this study are the lensing dispersion $\langle\kappa^2\rangle$ and the linear bias $b_{\mathrm{GW}}$, assuming that other parameters such as cosmological parameters and the linear bias of spectroscopic galaxies are well constrained by other observations. Errors in the lensing dispersion and the linear bias can be estimated from the Fisher information matrix components, using the auto-correlation  angular power spectrum $C^{\rm{ww}}(\ell)$ and the cross-correlation angular power spectrum $C^{\rm{wg}}(\ell)$. These constraints can be visually illustrated as confidence ellipses \citep{coe2009}.

\subsubsection{Joint Analysis}\label{subsubsec:joint analysis}
The constraining power on the parameters can be enhanced by combining the auto- and the cross-correlation angular power spectra, if the two observables have different dependence on the parameters. For instance, suppose the parameter dependence of the auto- and cross-correlation can be approximated as
\begin{align}
C^{\rm{ww}} \propto b_{\rm{GW}}^2 \langle\kappa^2\rangle^{\alpha}, \label{eq:alpha_taylor}\\
C^{\rm{wg}} \propto b_{\rm{GW}} \langle\kappa^2\rangle^{\beta}, \label{eq:beta_taylor}
\end{align}
where \( \alpha \) and \( \beta \) characterize the sensitivity of each observable to the lensing dispersion, the degeneracy directions between \( b_{\rm{GW}} \) and \( \langle\kappa^2\rangle \) differ for the two observables when $\alpha\neq2\beta$, allowing the degeneracy to be broken through their combination.  
This approach is referred to as a joint analysis. Since the auto- and cross-correlation are statistically correlated, adding the two Fisher matrices is not sufficient.  
The full covariance matrix, including the cross-covariance between \( C^{\rm{ww}} \) and \( C^{\rm{wg}} \), must be considered to correctly evaluate the joint Fisher matrix \citep{takada2007}. Our analysis explicitly incorporates the cross-covariance between the two angular power spectra into the Fisher matrix calculation to properly account for their statistical correlation.

\section{Result}\label{sec:result}
In this section, we show that the auto-correlation angular power spectrum of gravitational wave sources is a decreasing function of the lensing dispersion. This relationship enables us to estimate the lensing dispersion from the auto-correlation power spectrum of gravitational wave sources. We also show that the auto-correlation power spectrum can be measured with a sufficient signal-to-noise ratio. Furthermore, we demonstrate that the degeneracy between the lensing dispersion and the linear bias can be partially broken by combining the auto-correlation with the cross-correlation between binary black holes and spectroscopic galaxies. 

For gravitational wave observations, we consider third-generation detectors such as Einstein Telescope \citep{branchesi2023, maggiore2024}, Cosmic Explorer \citep{reitze2019a}, DECIGO \citep{tsuji2024a, kawamura2021}, and Big Bang Observer \citep{cutler2009a}. For spectroscopic galaxy surveys, we consider Euclid, which was launched in 2023 and is designed to observe galaxy clustering at high redshifts \citep{scaramella2022}.

The uncertainty in the luminosity distance of each gravitational wave sources is modeled as a log-normal distribution with a constant $\sigma_{\ln D}$, which represents the fractional distance error. For the sky localization, we take into account the finite angular resolution of the detectors by imposing a maximum multipole $\ell_{\max }$. 
We assume that the sky localization uncertainties of a significant number of gravitational wave sources in the era of third-generation detectors are smaller than the $\ell_{\rm{max}}$ chosen in our study. The effect of any correlation between errors on the luminosity distance and the sky localization is expected to be mitigated by adopting a conservative value of $\ell_{\mathrm{max}}$.


We adopt a redshift bin of \( z = [2.0, 2.3] \), corresponding to the range expected to be covered by Euclid. Unless otherwise stated, this redshift range is used throughout this paper. The luminosity distance bin is defined by the range of luminosity distances in a homogeneous and isotropic FLRW universe corresponding to the range of this redshift bin. The observational region on the celestial sphere is assumed to cover the entire sky, as gravitational wave detectors are sensitive to sources across the whole sky. While Euclid is expected to cover only about one-third of the sky, we ignore its effect in the present analysis for simplicity.

Following \citet{euclid_collaboration_euclid_2024}, we set a fiducial value of the linear bias of both binary black holes and spectroscopic galaxies to \( b_{\rm{GW}} \simeq b_{\rm{g}} \simeq 2.6 \). Although the linear bias is redshift-dependent, the effect of the redshift dependence of the linear bias is not significant, given the small width of the redshift bin.

The average number density of gravitational wave sources in Eq.~\eqref{eq:average number density} is given by
\begin{align}
\bar{n}_{\rm{GW}}(z) = T_{\mathrm{obs}} \frac{\dot{n}_{\mathrm{GW}}(z)}{1+z},
\end{align}
where \( T_{\rm{obs}} \) is the observation time and \( \dot{n}_{\mathrm{GW}}(z) \) denotes the binary black hole merger rate.  The observation time is set to \( T_{\rm{obs}} = 10\;\mathrm{yr} \), based on the planned operational periods of the Einstein Telescope.

The merger rate at \( z = 0.2 \) is estimated to be \( R_0 = 19\;\text{--}\;42\;\mathrm{Gpc}^{-3}\,\mathrm{yr}^{-1} \)\citep{the_ligo_scientific_collaboration_population_2023}. This estimate is based on the GWTC-3 catalog, which compiles data from the LIGO-Virgo joint observations (O1–O3). At redshifts \( z \lesssim 1 \), the merger rate evolves approximately as \( (1+z)^\kappa \), with \( \kappa = 2.9^{+1.7}_{-1.8} \) \citep{the_ligo_scientific_collaboration_population_2023}. While the merger rate of binary black holes at higher redshifts is uncertain, it is reasonable to assume that the merger rate keeps increasing with redshift at least out to $z\sim2$, assuming that the merger rate more or less traces the cosmic star formation rate density that increases out to $z\simeq 2$ (e.g., \citep{madau2014,behroozi2019,bouwens2020}). We therefore simply extrapolate this power-law evolution model. In this model, the merger rate at \( z \simeq 2 \) is estimated as
\begin{align}
\nonumber \dot{n}_{\rm{GW}}(z=2.0) &\sim R_0 \times \left( \frac{1+2}{1+0.2} \right)^3 \\
&\sim15.6\times R_0 .
\end{align}
In this analysis, we assume a binary black hole merger rate of  \( \dot{n}_{\mathrm{GW}} = 2\times 10^{-6}\;h^3\,\mathrm{Mpc}^{-3}\,\mathrm{yr}^{-1} \) at \( z = 2 \). This corresponds to \( R_0 \sim 40\;\mathrm{Gpc}^{-3}\,\mathrm{yr}^{-1} \), which lies at the higher end of the GWTC-3 range. Although there are considerable uncertainties in the merger rate estimation, this value falls within the 90\% confidence interval and thus serves as a reasonable assumption. For spectroscopic galaxies, we adopt a number density of \( \bar{n}_{\mathrm{g}} = 4 \times 10^{-4}\;h^3\,\mathrm{Mpc}^{-3} \), following \citet{euclid_collaboration_euclid_2024}.

\subsection{The Relationship Between the Angular Power Spectrum and the Lensing Dispersion}\label{Relationship C and kappa2}
\begin{figure}[t]
\centering
\includegraphics[width=0.45\textwidth]{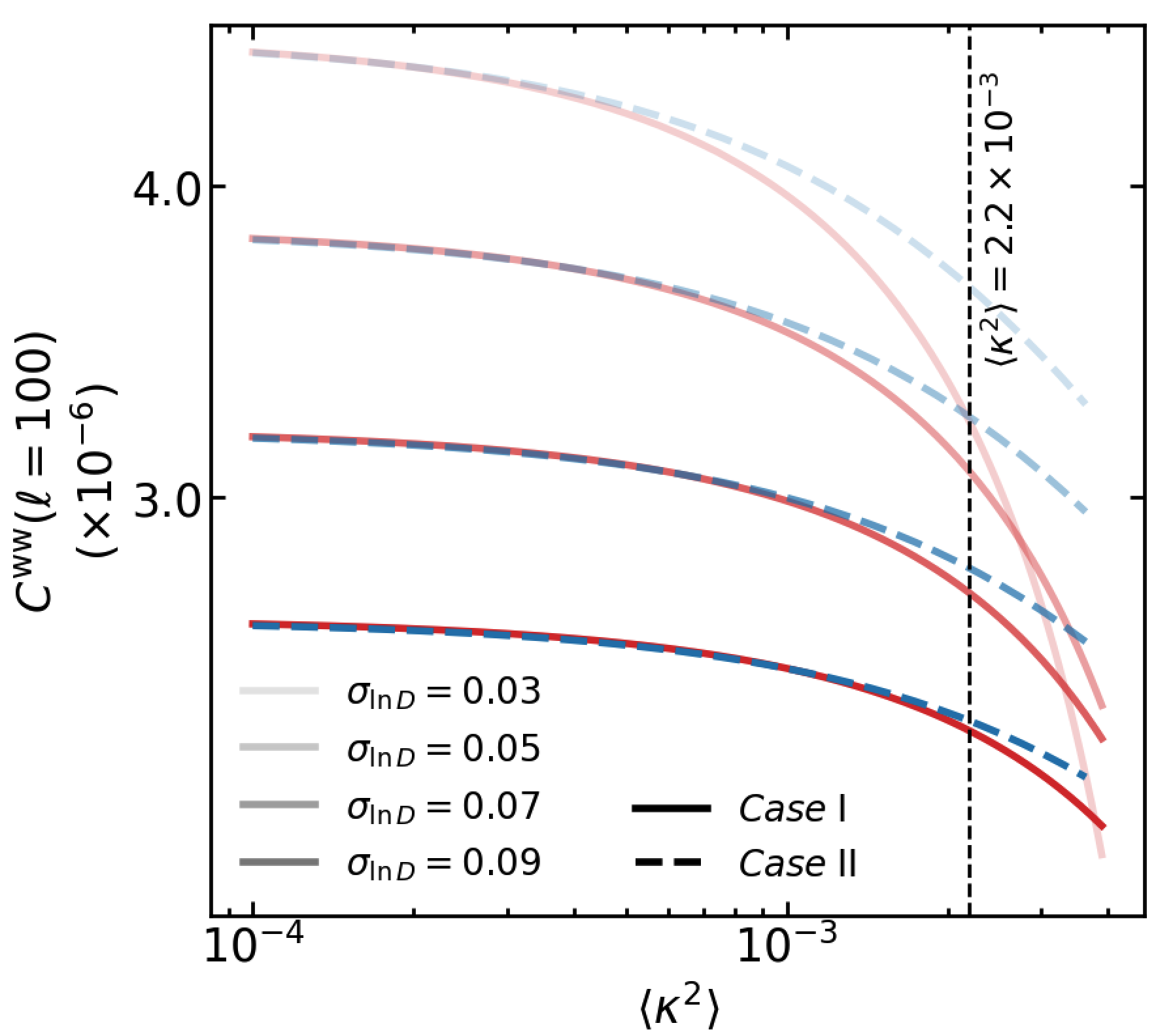}
\hfill
\includegraphics[width=0.45\textwidth]{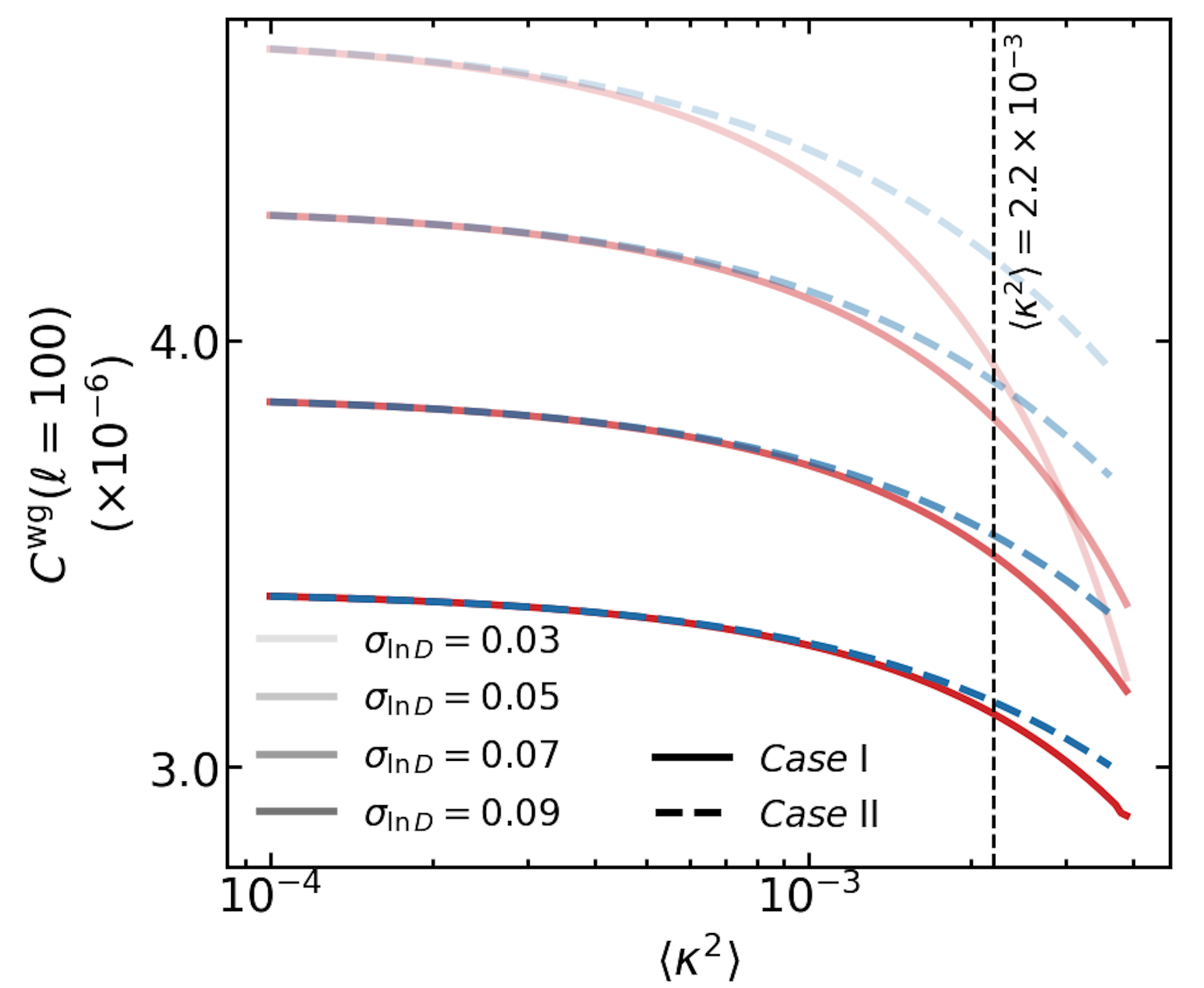}
\caption{The auto- (\textit {top}) and cross-correlation  (\textit{bottom})  angular power spectra at the multipole $\ell=100$ as a function of the lensing dispersion $\langle\kappa^2\rangle$. Each plot shows results for different observational error $\sigma_{\ln D}$, with solid lines representing Case$\;\rm{I}$ and dashed lines representing Case$\;\rm{I\hspace{-1.2pt}I}$.  
The vertical dashed line indicates a fiducial value of the lensing dispersion at $z=2.15$ taken from ray-tracing simulations
 \citep{takahashi2011}. }
\label{fig:Cww}
\end{figure}

Fig.~\ref{fig:Cww} shows the auto-correlation angular power spectrum of binary black holes and the cross-correlation angular power spectrum between binary black holes and spectroscopic galaxies, plotted as a function of the lensing dispersion. As shown in Fig.~\ref{fig:Cww}, both the angular power spectra decrease with increasing the lensing dispersion. This behavior can be understood as follows. The lensing effect alters the luminosity distances to the sources, and broadens out the distribution of gravitational wave sources along the line of sight for a given luminosity distance bin. The increased characteristic width of the radial distribution $\Delta \chi$ decreases the amplitude of both the auto- and cross-correlation angular power spectra, because they can be generally approximated as
\begin{align}
C^{\rm{ww}}(\ell) = \int \frac{d\chi}{\chi^2} W^2(\chi) P_{\rm{m}} \propto \frac{1}{\Delta\chi} P_{\rm{m}}\left(k = \frac{\ell+1/2}{\chi}, \chi\right),
\end{align}
which indicates that the amplitude is inversely proportional to $\Delta \chi$. Similarly, the larger observational error $\sigma_{\ln D}$ also broadens the radial distribution and decreases the amplitude, as is shown in Fig.~\ref{fig:Cww}.

Comparing Case$\;\rm{I}$ and Case$\;\rm{I\hspace{-1.2pt}I}$, at higher lensing dispersion values, the angular power spectra in Case$\;\rm{I}$ decrease faster than those in Case$\;\rm{I\hspace{-1.2pt}I}$. As discussed in Sec.~\ref{sec:cross-corelation}, this difference arises from the non-uniform broadening of the gravitational wave source distribution in Case$\;\rm{I}$.  We also observe that the solid lines in Fig.~\ref{fig:Cww} corresponding to Case$\;\rm{I}$ intersect at large lensing dispersion values across different observational errors.  
This intersection implies a breakdown of the approximation adopted in Case$\;\rm{I}$, whereas such an intersection is absent in Case$\;\rm{I\hspace{-1.2pt}I}$. The breakdown occurs because the Taylor expansion in Eq. \eqref{eq:xDobs} becomes inaccurate for smaller values of $\sigma_{\ln D}$.

\subsection{Signal-to-Noise Ratio}\label{subsec:r_snr}
\begin{figure}
\centering
\includegraphics[width=0.45\textwidth]{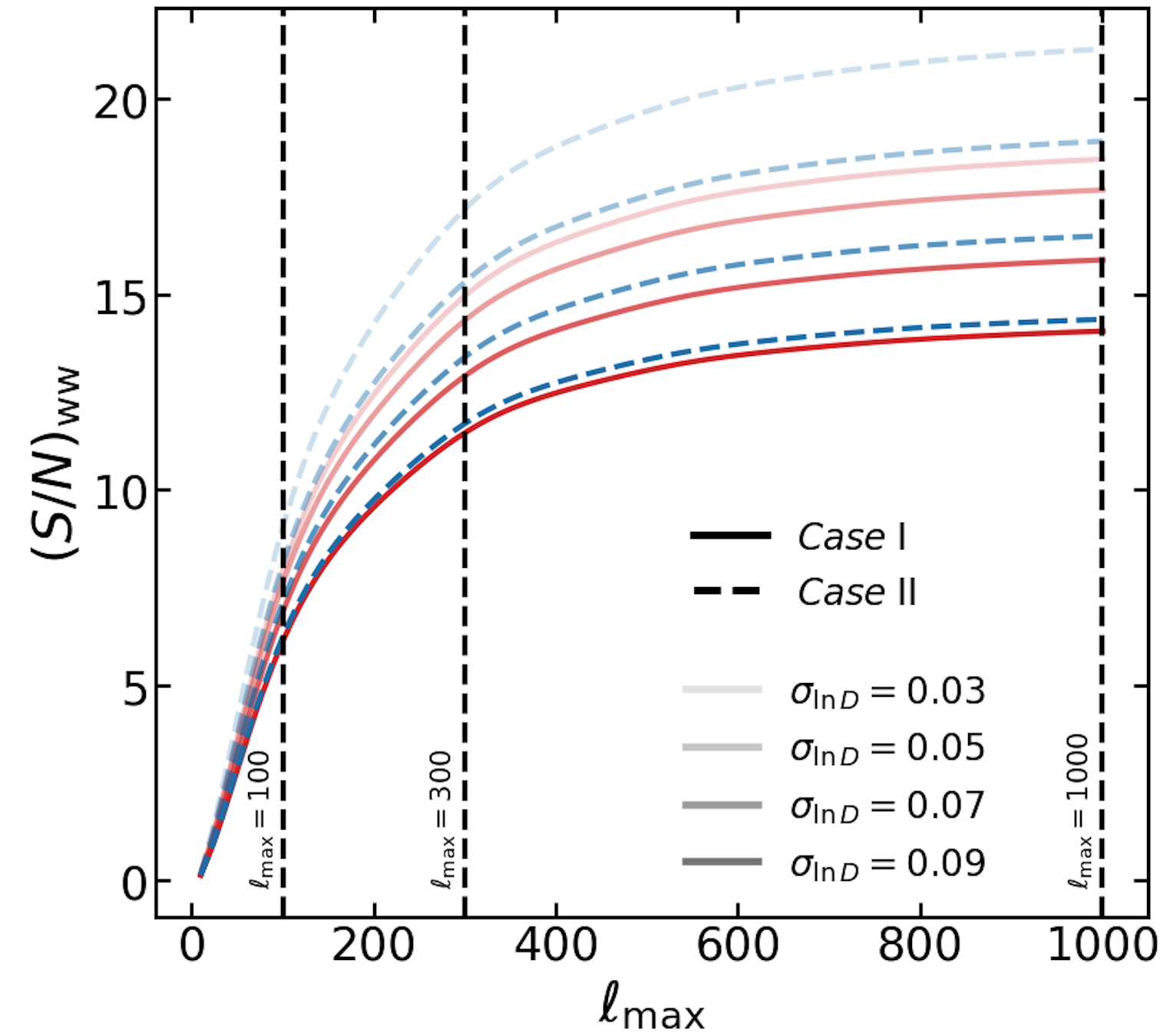}
\includegraphics[width=0.45\textwidth]{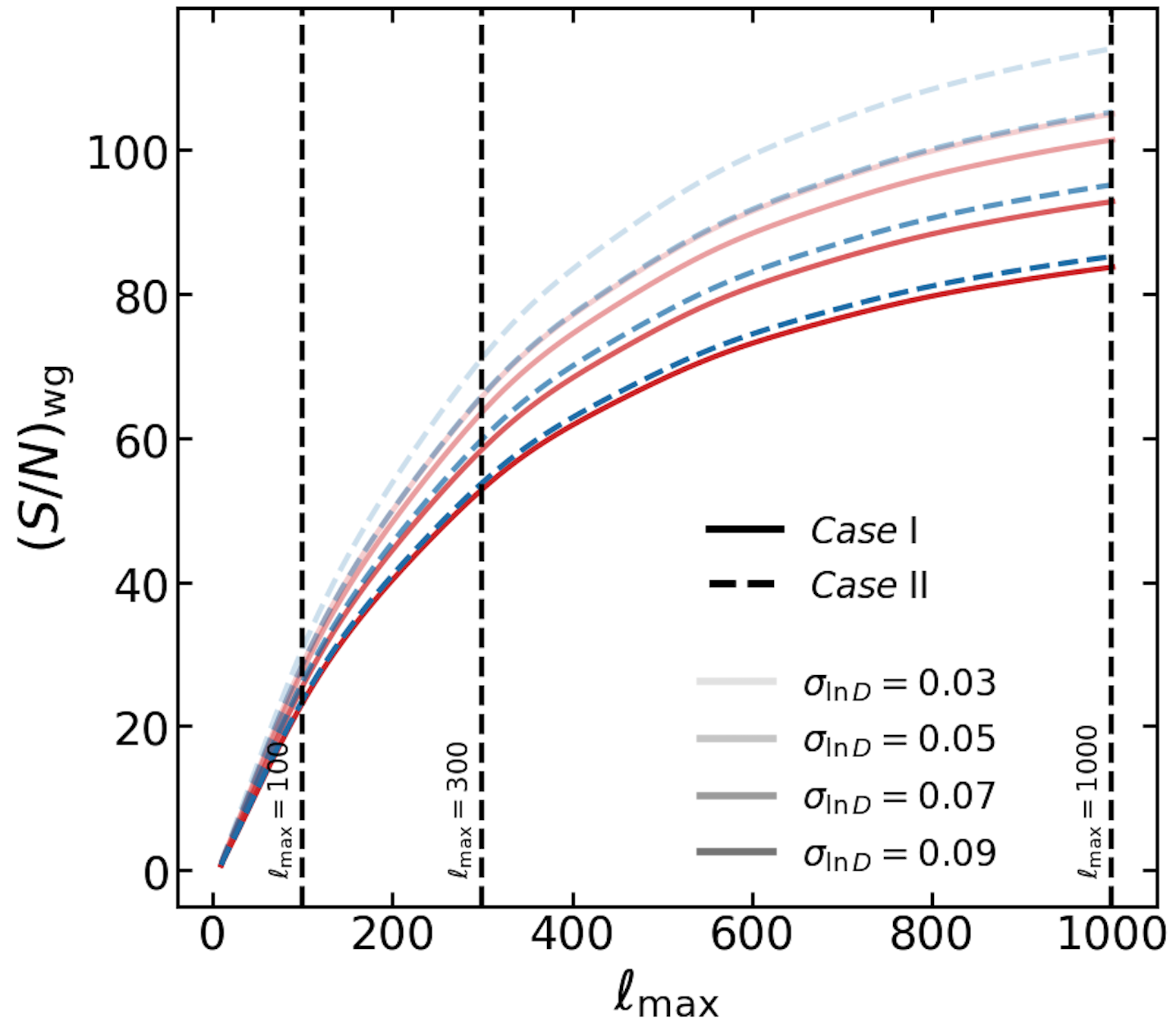}
\caption{The cumulative signal-to-noise ratios of the auto- (\textit{top}) and cross-correlation angular power spectra (\textit{bottom}). Each plot shows results for different values of the observational error $\sigma_{\ln D}$. Solid lines show results for  Case$\;\rm{I}$, and dashed lines for Case$\;\rm{I\hspace{-1.2pt}I}$. Vertical dashed lines indicate $\ell_{\rm{max}} = 100,\ 300$ and $1000$. The lensing dispersion is fixed at the fiducial value $\langle\kappa^2\rangle = 2.2 \times 10^{-3}$.}
\label{fig:snr}
\end{figure}

We assess whether the angular power spectra are observable by evaluating their signal-to-noise ratios.  Fig.~\ref{fig:snr} presents the cumulative signal-to-noise ratios for the auto- and cross-correlation angular power spectra. In computing the cumulative signal-to-noise ratios, we exclude multipoles below \( \ell = 10 \), as the Limber approximation becomes inaccurate at large angular scales.

As shown in Fig.~\ref{fig:snr}, we find that both the angular power spectra can be observed with the signal-to-noise ratios much larger than unity. The cumulative signal-to-noise ratios of the auto- and cross-correlation angular power spectra exhibit different dependence on $\ell_{\mathrm{max}}$, where $\ell_{\mathrm{max}}$ is essentially determined by the localization precision of gravitational wave observations, primarily due to the significant difference in number densities between gravitational wave sources and spectroscopic galaxies, as discussed in Sec.~\ref{subsubsec:snr}. For example, in Case$\;\rm{I}$ with \( \sigma_{\ln D} = 0.05 \), the number density of gravitational wave sources is \( \bar{n}^{\rm{w}} = 3.6 \times 10^4 \), whereas the number density of spectroscopic galaxies is \( \bar{n}^{\rm{g}} = 2.3 \times 10^6 \) for fiducial luminosity distance and redshift bins. The relatively small number density of gravitational wave sources leads to a significant shot noise, which dominates the error in the auto-correlation measurement, as the shot noise scales with the inverse of the number density. The cumulative signal-to-noise ratio for the auto-correlation increases with \( \ell_{\rm{max}} \) up to approximately $\ell_{\mathrm{max}}=300$. Beyond this scale, however, the improvement saturates due to the dominant contribution of the shot noise.  In contrast, the cross-correlation angular power spectrum benefits from the high number density of galaxies. As a result, the shot noise of gravitational wave sources has a smaller impact on the cumulative signal-to-noise ratio. 

While third-generation detectors can detect almost all gravitational waves from binary black holes in the redshift range of our interest with a signal-to-noise ratio above 8, only a fraction of these events may be localized with the precision of $1~\mathrm{deg}^2$ \citep{iacovelli2022,branchesi2023}. As expected, the $S/N$ scales approximately with the fraction $x$ of events that achieve the desired localization precision, because the shot noise increases as the number of sources decreases. Specifically, the cumulative signal-to-noise ratio for the auto-correlation $(S/N)_{\mathrm{ww}}$ is reduced by factors of 10, 100, and 1000 for $x=0.1, 0.01, 0.001$, respectively, while the cumulative signal-to-noise ratio for the cross-correlation $(S/N)_{\rm {wg}}$ is reduced by factors of $\sqrt{10}, 10$, and $\sqrt{1000}$ for the same values of $x$.

We note that the localization precision depends strongly on the detector network geometry. For instance, combining the Einstein Telescope with a triangular configuration with L-shaped interferometers such as the Cosmic Explorer can significantly improve localization \citep{iacovelli2022,branchesi2023,maggiore2024,giovanni2025a}. Future networks with multiple third-generation detectors may enable sub-degree localizations for a larger fraction of events, especially with long-term operation and joint observations. Therefore, although the scenario assuming full localization is optimistic, it is not too unrealistic in the context of planned detector developments.

\subsection{The Degeneracy between the lensing dispersion and the Linear Bias, and the Fisher Analysis}\label{subsec:r_degeneracy}

\begin{figure}[t]
\centering
\includegraphics[width=0.46\textwidth]{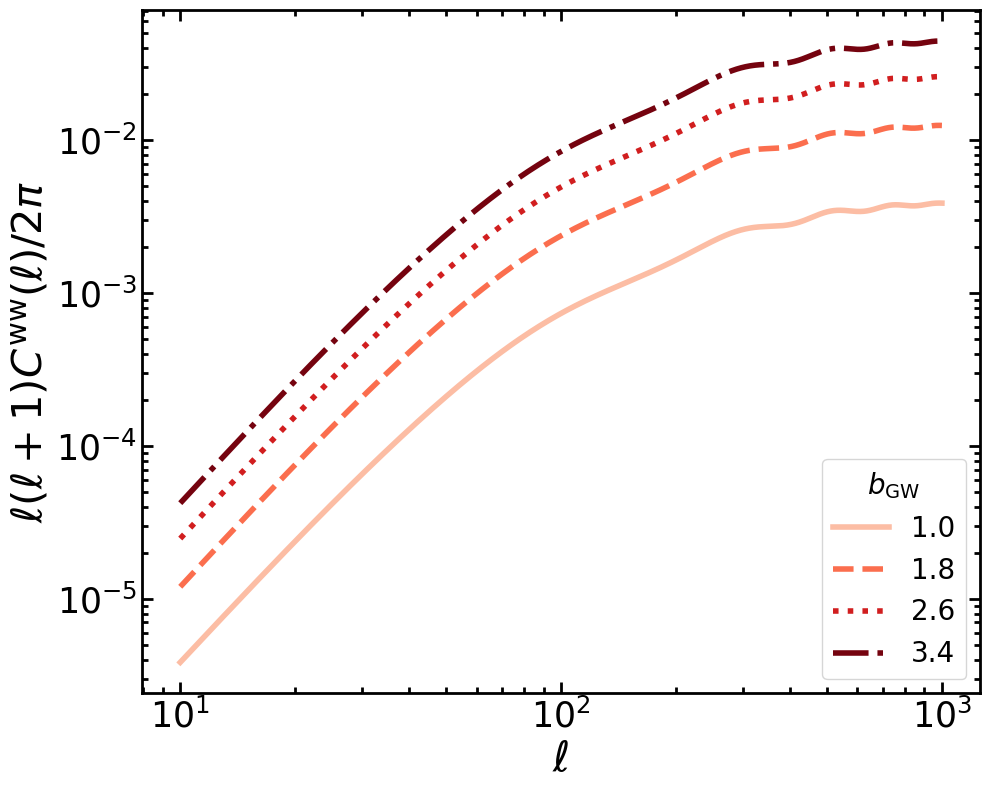}
\includegraphics[width=0.47\textwidth]{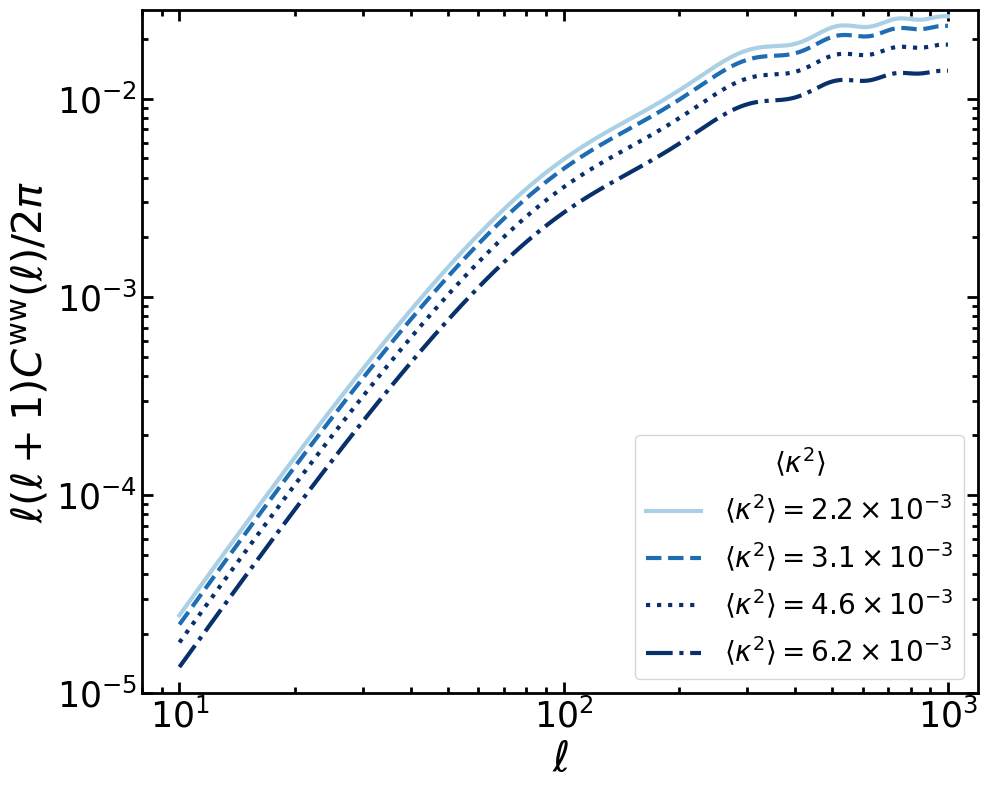}
\caption{The auto-correlation  angular power spectrum $C^{\rm{ww}}({\ell})$ for different values of the linear bias $b_{\mathrm{GW}}$ ({\it top}) and the lensing dispersion $\langle\kappa^2\rangle$ ({\it bottom}).}
\label{fig:degeneracy}
\end{figure}

In this section, we discuss a strong degeneracy between the lensing dispersion $\langle\kappa^2\rangle$ and the linear bias $ b_{\rm GW} $. We also demonstrate that a joint analysis of the auto- and cross-correlation angular power spectra can partially break this degeneracy.

Fig.~\ref{fig:degeneracy} shows the dependence of the auto-correlation angular power spectrum on the lensing dispersion and the linear bias. It is found that both  parameters shift the amplitudes, which immediately suggests that these two parameters are degenerate, even if we assume that cosmological parameters are well constrained from other observations.  A similar degeneracy also exists in the cross-correlation angular power spectrum. Therefore, neither the auto-correlation nor the cross-correlation alone can constrain the lensing dispersion precisely.
However, by numerically evaluating values of $\alpha$ and $\beta$ in Eq.~\eqref{eq:alpha_taylor} and~Eq.~\eqref{eq:beta_taylor}, we find
\begin{align}
\alpha\simeq -0.26,\quad\beta\simeq -0.15,
\end{align}
for Case I. Since $\alpha\neq2\beta$, we expect that the degeneracy can be partially broken by the joint analysis of the auto- and cross-correlation angular power spectra, as discussed in Sec.~\ref{subsubsec:joint analysis}.
\begin{figure}[t]
\centering
\includegraphics[width=0.45\textwidth]{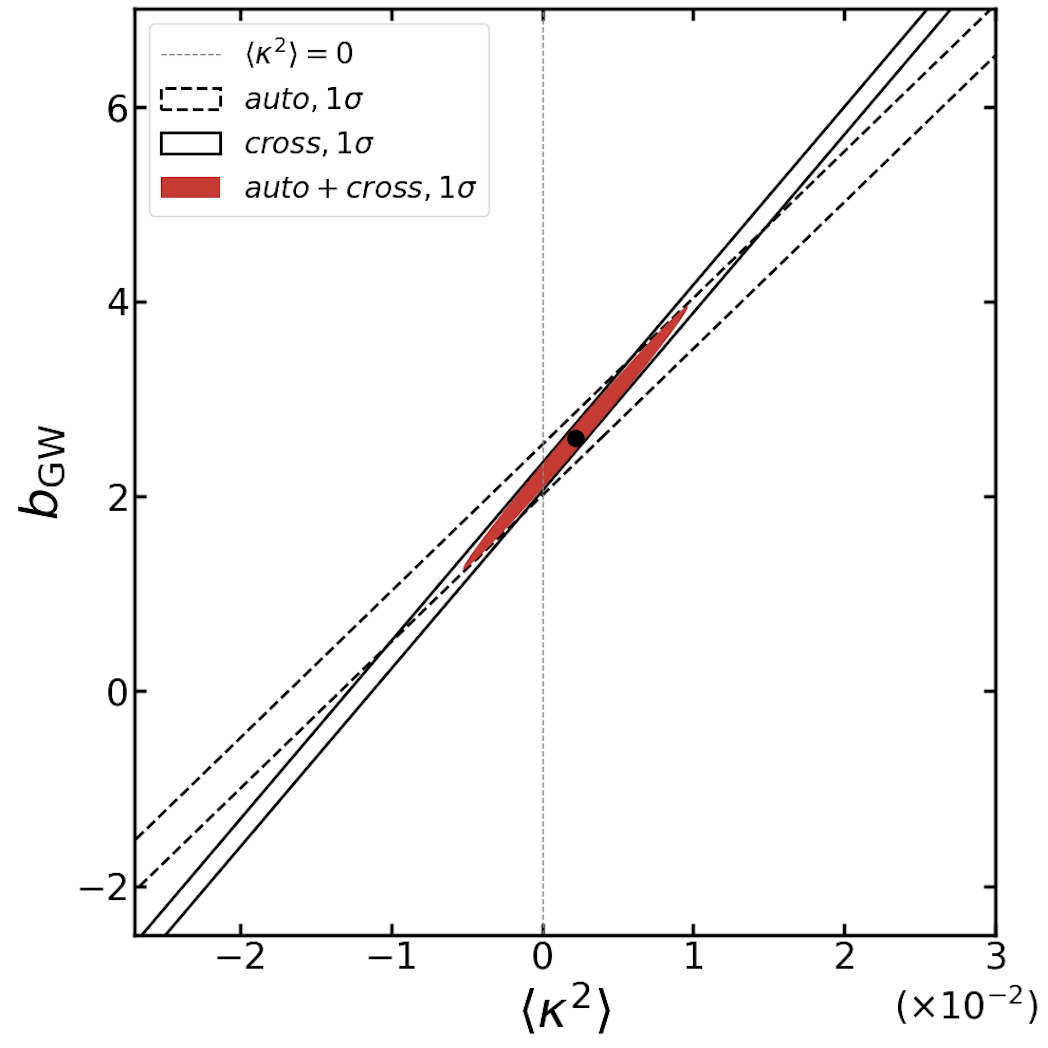}
\caption{Parameter constraints on the lensing dispersion $\left\langle\boldsymbol{\kappa}^2\right\rangle$ and the linear bias of binary black holes $b_{\mathrm{GW}}$ from the Fisher matrix analysis with the maximum multipole to $\ell_{\mathrm{max}}=100$ for Case I. The black dashed contour shows the $1 \sigma$ confidence region from the auto-correlation angular power spectrum, while the solid black contour is from the cross-correlation angular power spectrum. The red-shaded region represents the joint constraint from both the auto- and cross-correlations.}
\label{fig:fisher_auto_cross}
\end{figure}

\begin{figure}[t]
\label{fig:epsart} 
\centering
\includegraphics[width=0.45\textwidth]{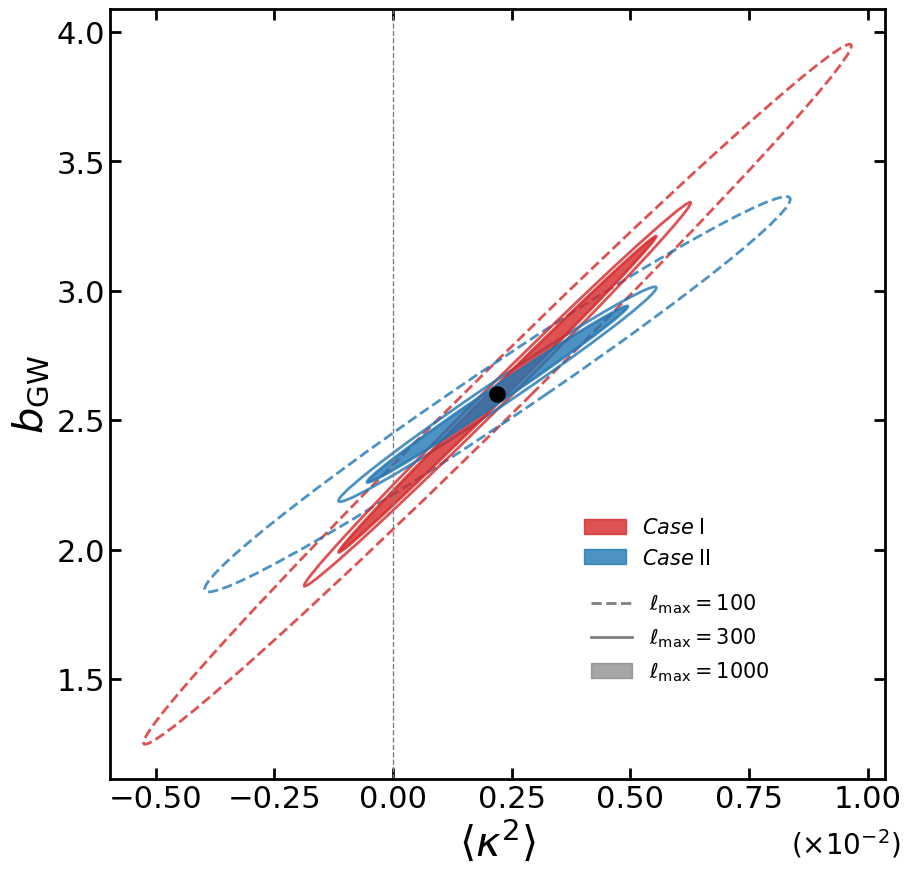}
\caption{\label{fig:epsart}Comparison of the $1 \sigma$ constraint contours for Case$\;\rm{I}$
(\textit{red}) and Case$\;\rm{I\hspace{-1.2pt}I}$ (\textit{blue)}, shown in the parameter space of the lensing dispersion $\left\langle\kappa^2\right\rangle$ and the linear bias $b_{\mathrm{GW}}$ of gravitational wave sources. Different contours show results for different maximum multipoles $\ell_{\mathrm{max}}=100$ ({\it dashed}), $300$ ({\it solid}), and $1000$ ({\it shaded}), reflecting different localization precisions of binary black holes, included in the analysis.}
  \label{fig:fisher_case1_case2}
\end{figure}

\renewcommand{\arraystretch}{1.6} 
\setlength{\tabcolsep}{12pt}
\begin{table*}[t]
\centering
\begin{tabular}{lcccc}
\hline \hline
\multicolumn{1}{c}{\textbf{Case$\;\rm{I}$}} &
$\sigma_{\langle\kappa^2\rangle}$ &
$\sigma_{\langle\kappa^2\rangle}(b_{\rm{GW}}=2.6)$ &
$\sigma_{b_{\rm{GW}}}$ &
$\sigma_{b_{\rm{GW}}}(\langle\kappa^2\rangle=2.2\times10^{-3})$\\
\hline
$\ell_{\max }=100$ & $4.9\times\;10^{-3}$&$4.8\times\;10^{-4}$& 0.89&$8.6\times10^{-2}$\\
$\ell_{\max }=300$ & $2.7\times\;10^{-3}$&$2.1\times\;10^{-4}$& 0.49&$3.9\times10^{-2}$\\
$\ell_{\max }=1000$ & $2.2\times\;10^{-3}$&$1.4\times\;10^{-4}$& 0.40&$2.5\times10^{-2}$\\
\hline
\quad\textbf{Case$\;\rm{I\hspace{-1.2pt}I}$} &
$\sigma_{\langle\kappa^2\rangle}$ &
$\sigma_{\langle\kappa^2\rangle}(b_{\rm{GW}}=2.6)$ &
$\sigma_{b_{\rm{GW}}}$ &
$\sigma_{b_{\rm{GW}}}(\langle\kappa^2\rangle=2.2\times10^{-3})$\\
\hline
$\ell_{\max }=100$ & $4.1\times\;10^{-2}$&$6.7 \times\;10^{-4}$& 0.50&$8.4\times10^{-2}$\\
$\ell_{\max }=300$ & $2.2\times\;10^{-2}$&$3.0\times\;10^{-4}$& 0.27&$2.7\times10^{-2}$\\
$\ell_{\max }=1000$ & $1.8\times\;10^{-2}$&$1.9\times\;10^{-4}$&0.22&$2.2\times10^{-2}$\\
\hline \hline
\end{tabular}
\caption{Expected $1\sigma$ errors for the lensing dispersion $\langle\kappa^2\rangle$ and the linear bias $b_{\rm{GW}}$ of gravitational wave sources for Case$\;\rm{I}$ and Case$\;\rm{I\hspace{-1.2pt}I}$. Each row corresponds to a different maximum multipole $\ell_{\rm{max}}$. From left to right, each column shows the error on $\langle\kappa^2\rangle$ marginalized over $b_{\rm{GW}}$, the error on $\langle\kappa^2\rangle$ with $b_{\rm{GW}}$ fixed at $2.6$, the error on $b_{\rm{GW}}$ marginalized over $\langle\kappa^2\rangle$, and the error on $b_{\rm{GW}}$ with $\langle\kappa^2\rangle$ fixed at $2.2 \times 10^{-3}$. 
}
\label{tab:marginal_error}
\end{table*}

Fig.~\ref{fig:fisher_auto_cross} shows the \(1\sigma\) ($68.3\%$) confidence ellipses obtained from auto- and cross-correlation analyses for \(\ell_{\rm max}=100\), which clearly shows the advantage of a joint analysis. The auto- or cross-correlation analysis results in a significant degeneracy between the linear bias \(b_{\rm GW}\) and the lensing dispersion \(\langle\kappa^2\rangle\), making the detection of \(\langle\kappa^2\rangle\) almost impossible. However,  combining both auto- and cross-correlation measurements partially breaks this degeneracy. 

Table~\ref{tab:marginal_error} summarizes the expected \(1\sigma\) uncertainties on the lensing dispersion \(\langle\kappa^2\rangle\) and the linear bias \(b_{\rm GW}\) for both Case$\;\rm{I}$ and Case$\;\rm{I\hspace{-1.2pt}I}$. Due to the degeneracy between the linear bias and the lensing dispersion, marginalized constraints on these parameters are significantly degraded compared to the case where the other parameter is fixed. The results also clearly show that increasing \(\ell_{\rm max}\) tightens the constraints, highlighting the importance of good localization precision for gravitational wave sources. In particular, increasing  \(\ell_{\rm max}\) from \(\ell_{\rm max}=100\) to \(\ell_{\rm max}=300\) significantly improves the parameter constraints.

Fig.~\ref{fig:fisher_case1_case2} illustrates the \(1\sigma\) confidence ellipses obtained from joint analyses for \(\ell_{\rm max}=100, 300,\) and \(1000\). The confidence ellipse includes \(\langle\kappa^2\rangle=0\), indicating that the lensing dispersion is not significantly detected. Even in that case, it allows us to place an upper limit on the lensing dispersion. The difference between Case I and Case$\;\rm{I\hspace{-1.2pt}I}$ will be discussed in detail in Sec. \ref{subsec:systematic errors}.

\section{Discussion}\label{sec:discussion}
\subsection{Effect of the Merger Rate on Parameter Constraints}\label{d_merger rate}
\begin{figure}[t]
\centering
\includegraphics[width=0.45\textwidth]{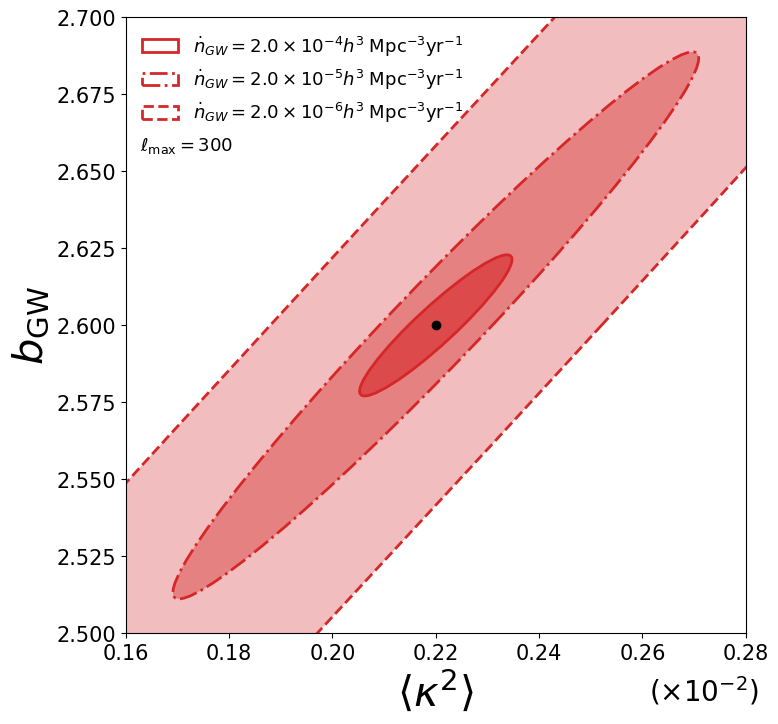}
\caption{\label{fig:epsart}Confidence ellipses in the parameter space of the lensing dispersion $\left\langle\kappa^2\right\rangle$ and the linear bias $b_{\mathrm{GW}}$ of gravitational wave sources, shown for three different merger rates. The solid line and darkest region correspond to $\dot{n}_{\mathrm{GW}}=2 \times10^{-4} h^3\;\mathrm{Mpc}^{-3} \mathrm{yr}^{-1}$, the dash-dotted line and medium-shaded region to $\dot{n}_{\mathrm{GW}}=2 \times10^{-5} h^3\;\mathrm{Mpc}^{-3} \mathrm{yr}^{-1}$, and the dashed line and lightest region to $\dot{n}_{\mathrm{GW}}=2 \times10^{-6} h^3\;\mathrm{Mpc}^{-3} \mathrm{yr}^{-1}$. The maximum multiple is $\ell_{\mathrm{max}}=300$. }
\label{fig:R_upper}
\end{figure}

The merger rate of gravitational wave sources, which is not yet tightly constrained at high redshifts in observations, significantly affects the strength of the parameter constraints. Fig.~\ref{fig:R_upper} shows confidence ellipses for the baseline merger rate $\dot{n}_{\mathrm{GW}}=2\times 10^{-6}h^3\;\mathrm{Mpc}^{-3}\mathrm{yr}^{-1}$ as well as the higher merger rates of $2\times 10^{-5}h^3\;\mathrm{Mpc}^{-3}\mathrm{yr}^{-1}$ and $2\times 10^{-4}h^3\;\mathrm{Mpc}^{-3}\mathrm{yr}^{-1}$, calculated using Case$\;\mathrm{I}$. As the merger rate increases, the number density of gravitational wave sources also increases. This leads to a reduction in the shot noise and, consequently, tighter parameter constraints. A similar improvement can be achieved by extending the observation time to accumulate more events. This has important implications for future detectors such as DECIGO or Einstein Telescope, which are expected to operate over long durations and detect large numbers of compact binary mergers. We note, however, that the highest merger rate considered here $2 \times 10^{-4} h^3~\mathrm{Mpc}^{-3} \mathrm{yr}^{-1}$ is highly optimistic, as it is comparable to the number density of galaxies and would imply nearly one merger per galaxy per year at $z \sim 2$. In practice, such an effective merger rate may be achieved by combining multiple types of compact binary mergers e.g., binary neutron stars and black hole-neutron star binaries and by longer observation time.

\subsection{Effect of the Luminosity Distance and Redshift Bin Widths}\label{subsec:d_redshift bin width}
\begin{figure}[t]
\centering
\includegraphics[width=0.45\textwidth]{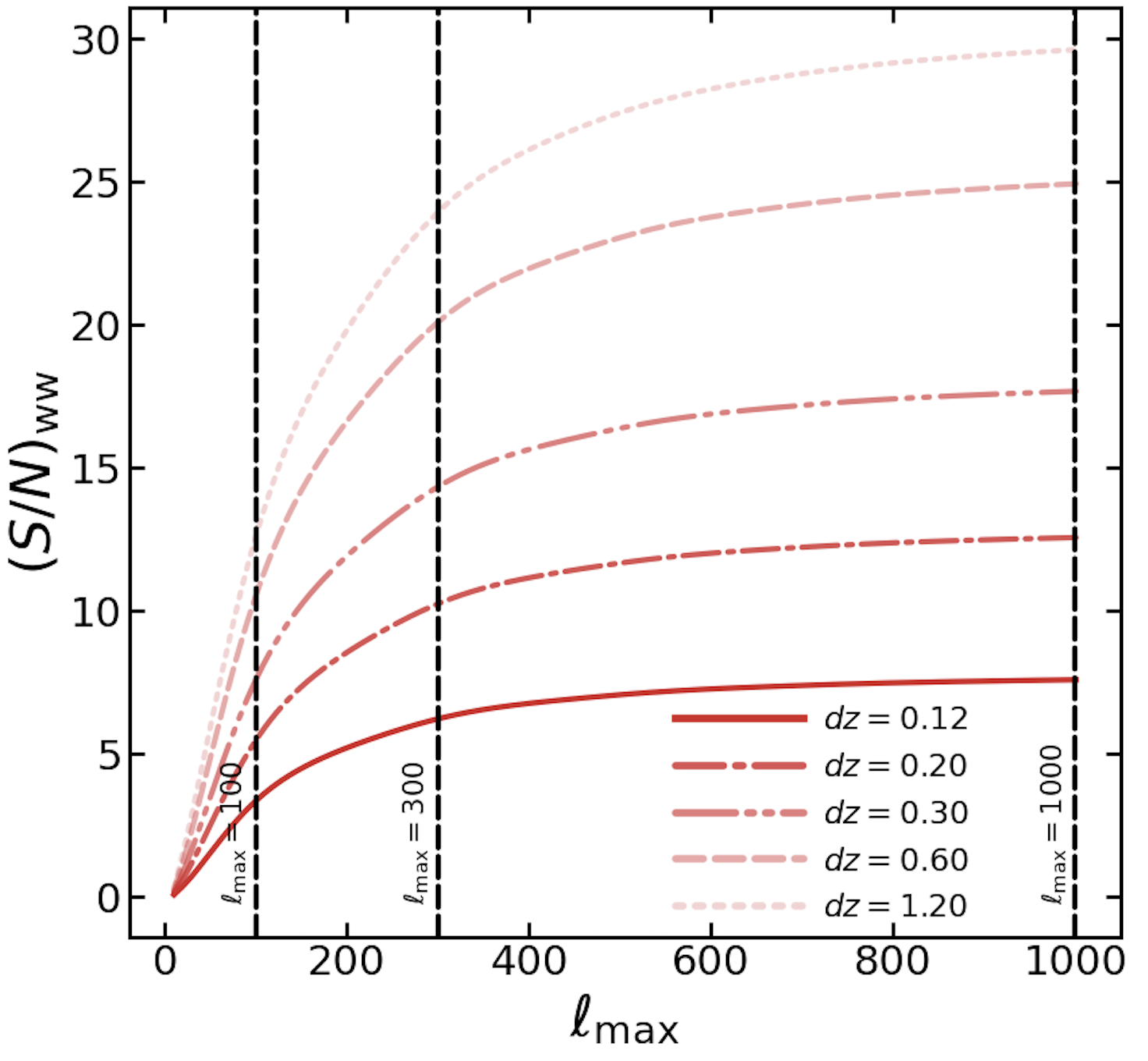}
\hfill
\includegraphics[width=0.45\textwidth]{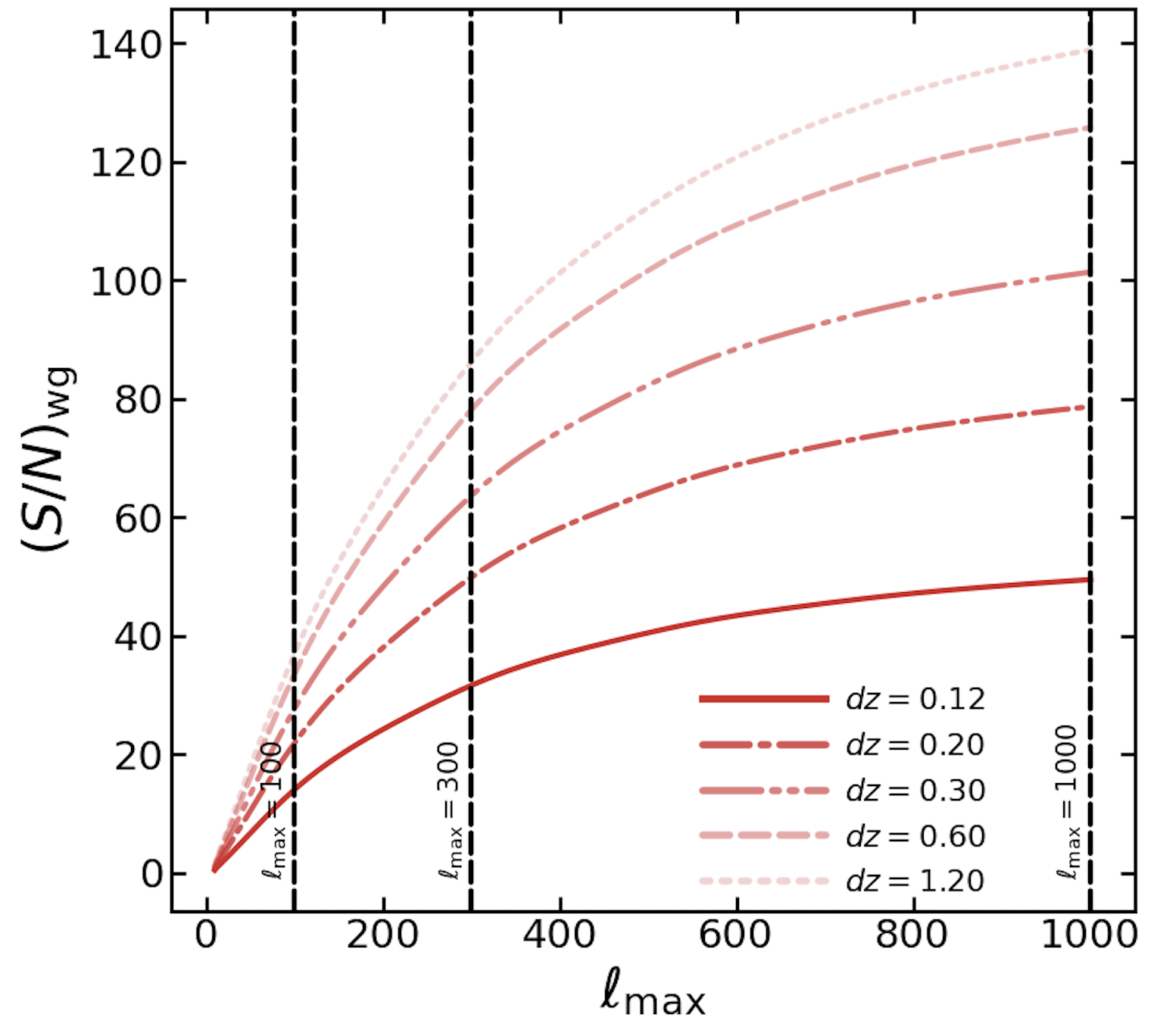}
\caption{The signal-to-noise ratio for varying redshift and luminosity distance bin widths for the auto-({\it top}) and cross-correlation ({\it bottom}) angular power spectra. Redshift bin widths are $dz=0.12,0.2,0.3,0.6,$ and $1.2$. }
\label{fig:snr_dz}
\end{figure}

\begin{figure}[t]
\centering
\includegraphics[width=0.45\textwidth]{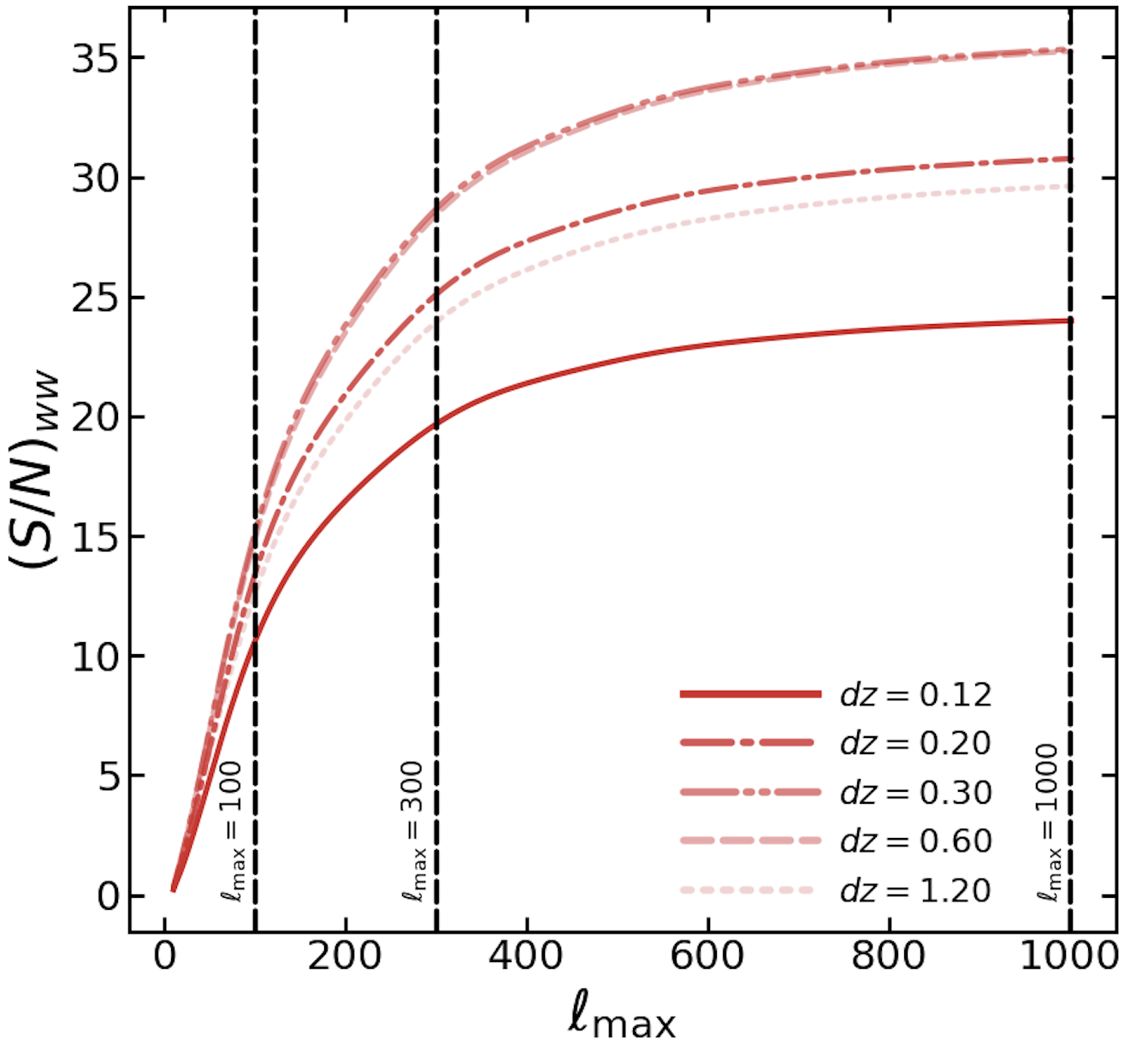}
\hfill
\includegraphics[width=0.45\textwidth]{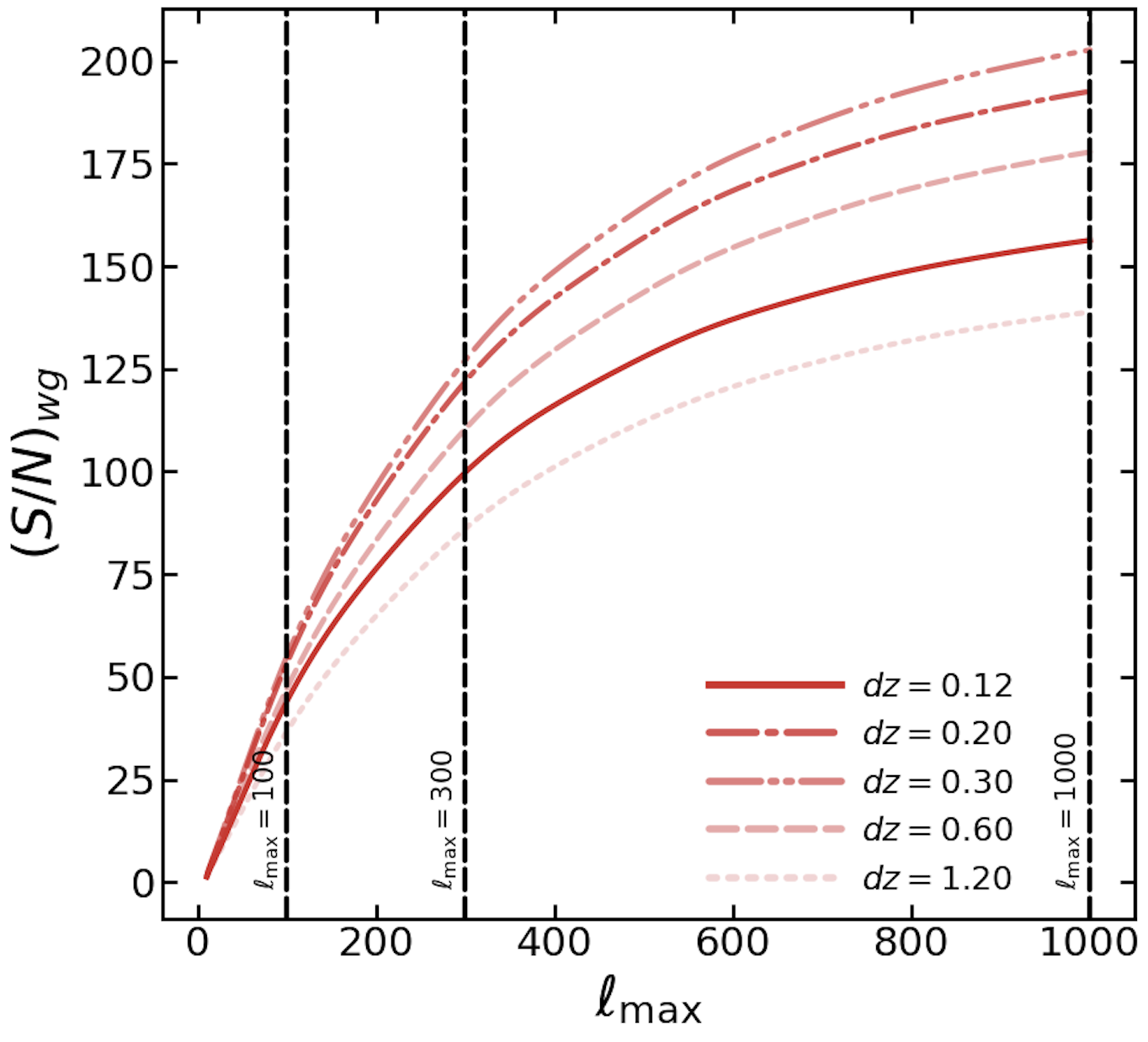}
\caption{Total signal-to-noise ratio obtained by summing the signal-to-noise ratios calculated for multiple redshift bins of width $dz$, keeping the total redshift range fixed to $1.2$. The luminosity distance bins are divided similarly to the redshift bins. The redshift bin widths $dz = 0.12,0.2,0.3,0.6$, and $1.2$ correspond to dividing the whole redshift range into $N = 10,6,4,2$, and $1$ bins, respectively.}
\label{fig:snr_div}
\end{figure}

We investigate the influence of the widths of the luminosity distance and redshift bins on the signal-to-noise ratio. In this analysis, we assume the constant black hole merger rate of \( \dot{n}_{\mathrm{GW}} = 2\times 10^{-6}h^3~\mathrm{Mpc}^{-3}\,\mathrm{yr}^{-1} \) for all redshift bins. Fig.~\ref{fig:snr_dz} shows the signal-to-noise ratio for different redshift bin widths (\( dz = 0.12, 0.2, 0.3, 0.6, 1.2 \)) and equivalent luminosity distance bin widths, computed using Case$\;\mathrm{I}$. We find that narrower bins reduce the number density of gravitational wave sources, increasing the shot noise, and reducing the signal-to-noise ratio. 

Next, we fix the total redshift range to \( dz = 1.2 \) and divide it into multiple smaller bins to investigate the variation of the signal-to-noise ratio. This cumulative signal-to-noise ratio is given by
\begin{align}
    S/N=\sqrt{\sum^N_i(S/N)^2_i}\;,
\end{align}
where $(S/N)_i$ denotes the signal-to-noise ratio computed for the $i$-th sub-bin after dividing the total redshift bin into $N$ sub-bins. As shown in Fig.~\ref{fig:snr_div}, subdividing the whole range into two to four sub bins increases the cumulative signal-to-noise ratio. In general, a larger number of sub-bins $N$ and equivalently a smaller redshift bin width $dz$ leads to enhances the signal amplitude in each bin, potentially resulting in a higher cumulative signal-to-noise ratio. However, increasing the number of sub-bins further does not improve the cumulative signal-to-noise ratio, as the signal-to-noise ratio in each sub-bin becomes too small to contribute significantly (see also Fig.~\ref{fig:snr_dz}). In particular, the case with $dz=1.2$ shows a relatively high cumulative signal-to-noise ratio for the auto-correlation $(S / N)_{\mathrm{ww}}$, even though it has a broader redshift range and thus a more diluted signal. The improvement in the cumulative signal-to-noise ratio is at most approximately $20\%$ for the auto-correlation, while it reaches up to $70\%$ for the cross-correlation for $\ell_{\rm{max}}=300$. This implies that a sufficiently high number density of gravitational wave sources is required to increase the cumulative signal-to-noise ratio by dividing the data into multiple smaller redshift bins and summing them up.

\subsection{Difference Between Case$\;\mathrm{I}$ and Case$\;\mathrm{I\hspace{-1.2pt}I}$}\label{subsec:systematic errors}
We use two approximation methods (Case$\;\mathrm{I}$ and Case$\;\mathrm{I\hspace{-1.2pt}I}$) for the gravitational lensing effect on luminosity distances to assess their impact on parameter constraints for the lensing dispersion and the linear bias. As shown in Fig.~\ref{fig:Cww}, the two methods agree well when the lensing dispersion is small. However, the difference between the methods becomes significant as the lensing dispersion increases. Fig.~\ref{fig:fisher_case1_case2} presents the 1$\sigma$ confidence ellipses for both methods. The constraints obtained from Case$\;\rm{I\hspace{-1.2pt}I}$ are tighter and exhibit less degeneracy compared to those from Case~$\rm{I}$, which can be attributed to systematic differences between these two approximation methods.

Since both Case$~\rm{I}$ and Case$\;\rm{I\hspace{-1.2pt}I}$ are based on approximations, neither can be considered better, but each has its characteristics. Case~$\rm{I}$ employs a second-order Taylor expansion of the lensing convergence, allowing for a detailed examination of the gravitational lensing effects on the distribution of gravitational wave sources. However, this approximation breaks down when the lensing dispersion becomes large or when observational uncertainties are small. In contrast, Case$\;\rm{I\hspace{-1.2pt}I}$ assumes a log-normal distribution for the lensing magnification. It is computationally more efficient and remains stable even for large values of the lensing dispersion. However, its limitation lies in its inability to describe the detailed structure of the lensing convergence distribution. Since the lensing dispersion increases with redshift \citep{takahashi2011}, these methodological differences may become increasingly important in high-redshift or high-dispersion regimes. We expect that the differences in results between these two methods provides a rough estimate of systematic errors due to approximations inherent to these methods.

\subsection{Breaking the Degeneracy Between the Lensing Dispersion and the Linear Bias}

It has been argued that measurements of the linear bias of binary black holes may provide a helpful clue to its origin (e.g., \citep{namikawa2016b,raccanelli2016,scelfo2018,calore2020,dehghani2025}). Recently, several studies have focused on how auto- and cross-correlations constrain the clustering bias of binary black holes to high redshifts (e.g., \citep{mukherjee2021, zazzera2024, zazzera2025}). Our results indicate that the degeneracy of the linear bias with the lensing dispersion should be appropriately considered when analyzing the angular clustering data to measure the linear bias. 

Here we discuss the possibility of breaking the degeneracy between the lensing dispersion and the linear bias of binary black holes. One approach is to make use of the scale-dependence of the bias parameter $b_{\mathrm{GW}}$. While in our analysis we assume the linear bias for simplicity, the scale dependence of the bias parameter can in principle arise at small scales or equivalently high $\ell$. Since our analysis indicates that the effect of the lensing dispersion is approximately scale-independent, the scale-dependence of the bias parameter could break the degeneracy, if the scale-dependence is well understood.
Another approach is to constrain the formation mechanisms of binary black holes from low redshift observations and predict their linear bias at higher redshifts. The formation mechanisms of binary black holes remain poorly understood, with multiple competing scenarios proposed in the literature (e.g., \citep{collaboration2019a,mapelli2020}). However, at low redshifts, we can measure the luminosity distance and the sky localization of binary black holes with higher precision, making it possible to identify their host galaxies in some cases. This enables us to investigate the environments in which binary black holes form and to constrain their formation channels (e.g., \citep{lamberts2016,vijaykumar2023,vijaykumar2024a,srinivasan2023}). In addition, at low redshifts the effect of the lensing dispersion on the angular clustering is negligible, which indicates that the linear bias of gravitational wave sources can be constrained from the low-redshift angular clustering analysis independently of the lensing dispersion. By imposing the value of the linear bias at high redshifts that is extrapolated from low redshifts  (e.g., \citep{toffano2019,neijssel2019,artale2020,santoliquido2022,peron2024}) as a prior, one can improve constraints on the lensing dispersion from the angular clustering analysis at high redshifts.

\section{Conclusion}\label{sec:conclusion}
The lensing dispersion contains rich cosmological information and is a key quantity for advancing our understanding of the small-scale structure of the Universe. At high redshifts, gravitational waves have the potential to serve as a powerful tool for measuring the lensing dispersion. However, the lack of redshift information limits its effectiveness. Developing new methods to measure the lensing dispersion without relying on redshift information is crucial for advancing the applications of gravitational waves at high redshifts.

In this paper, we have developed a method to measure the lensing dispersion using gravitational wave sources without requiring redshift data. Our approach is to utilize the angular clustering of gravitational wave sources to constrain the dispersion of the distance-redshift relation. We have found that the amplitudes of the angular power spectra of gravitational wave sources are a decreasing function of the lensing dispersion. Assuming that a sufficient number of gravitational wave events can be localized with angular uncertainties smaller than the scale corresponding to $\ell_{\mathrm{max}}=100$, we have confirmed that the angular power spectra can be measured with reasonable signal-to-noise ratios. However, there is a significant degeneracy between the lensing dispersion and the linear bias of gravitational wave sources. Based on the Fisher analysis, we have shown that a joint analysis combining the auto-correlation of gravitational wave sources with their cross-correlation with spectroscopic galaxies partially breaks the degeneracy to place meaningful constraints on the lensing dispersion. 

\begin{acknowledgments}
We would like to thank Atsushi Nishizawa, Ken Osato and Takahiro Nishimichi for useful discussions. We thank the anonymous referee for constructive comments, which led to a more realistic description of the merger rate based on the expected performances of third-generation gravitational wave detectors, as well as for suggestions that improved the clarity of the figures and deepened the discussion on the localization accuracy of gravitational-wave sources. This work was supported by JST SPRING, Grant Number JPMJSP2109. This work was supported by JSPS KAKENHI Grant Numbers JP25H00662, JP22K21349, and JP24K00684.

\end{acknowledgments}





%
\bibliographystyle{apsrev}
\bibliography{ref}
\end{document}